\documentclass[sigconf]{acmart}
\usepackage{multirow}
\AtBeginDocument{%
  \providecommand\BibTeX{{%
    \normalfont B\kern-0.5em{\scshape i\kern-0.25em b}\kern-0.8em\TeX}}}

\copyrightyear{2023}
\acmYear{2023}
\setcopyright{acmlicensed}\acmConference[KDD '23]{Proceedings of the 29th ACM SIGKDD Conference on Knowledge Discovery and Data Mining}{August 6--10, 2023}{Long Beach, CA, USA}
\acmBooktitle{Proceedings of the 29th ACM SIGKDD Conference on Knowledge Discovery and Data Mining (KDD '23), August 6--10, 2023, Long Beach, CA, USA}
\acmPrice{15.00}
\acmDOI{10.1145/3580305.3599484}
\acmISBN{979-8-4007-0103-0/23/08}

\usepackage{xcolor}
\usepackage[showdeletions]{color-edits}
\addauthor{il}{purple}
\addauthor{ju}{red}
\addauthor{aa}{blue}
\newcommand{\xhdr}[1]{\vspace{1.0mm}\noindent{{\bf #1.}}\hspace{0.5mm}}

\newtheorem*{fact*}{Fact}
\newtheorem{theorem}{Theorem}

\usepackage{caption}
\usepackage{subcaption}
\graphicspath{{images/}}
\acmSubmissionID{550}

\renewcommand\footnotetextcopyrightpermission[1]{}

\begin{document}

\title{Rank-heterogeneous preference models for school choice}

\author{Amel Awadelkarim}
\email{ameloa@stanford.edu}
\affiliation{%
  \institution{Stanford University}
  \city{Stanford}
  \state{CA}
  \country{USA}
  \postcode{94305}
}
\author{Arjun Seshadri}
\email{sesarjun@amazon.com}
\affiliation{%
  \institution{Amazon}
  \city{San Francisco}
  \state{CA}
  \country{USA}
  \postcode{}
}
\author{Itai Ashlagi}
\email{iashlagi@stanford.edu}
\affiliation{%
  \institution{Stanford University}
  \city{Stanford}
  \state{CA}
  \country{USA}
  \postcode{94305}
}
\author{Irene Lo}
\email{ilo@stanford.edu}
\affiliation{%
  \institution{Stanford University}
  \city{Stanford}
  \state{CA}
  \country{USA}
  \postcode{94305}
}
\author{Johan Ugander}
\email{jugander@stanford.edu}
\affiliation{%
  \institution{Stanford University}
  \city{Stanford}
  \state{CA}
  \country{USA}
  \postcode{94305}
}

\renewcommand{\shortauthors}{Awadelkarim et al.}

\begin{abstract}
School choice mechanism designers use discrete choice models to understand and predict families' preferences.
The most widely-used choice model, the multinomial logit (MNL), is linear in school and/or household attributes. While the model is simple and interpretable, it assumes the ranked preference lists arise from a choice process that is uniform throughout the ranking, from top to bottom. 
In this work, we introduce two strategies for rank-heterogeneous choice modeling tailored for school choice. First, we adapt a context-dependent random utility model (CDM), considering down-rank choices as occurring in the context of earlier up-rank choices. Second, we consider stratifying the choice modeling by rank, regularizing rank-adjacent models towards one another when appropriate. Using data on household preferences from the San Francisco Unified School District (SFUSD) across multiple years, we show that the contextual models considerably improve our out-of-sample evaluation metrics across all rank positions over the non-contextual models in the literature. Meanwhile, stratifying the model by rank can yield more accurate first-choice predictions while down-rank predictions are relatively unimproved. These models provide performance upgrades that school choice researchers can adopt to improve predictions and counterfactual analyses.
\end{abstract}






\maketitle

\section{Introduction}
Large school districts around the world employ \emph{school choice} mechanisms to assign students to K--12 schools. 
In many of these systems, families submit ranked preference lists over school programs to their district, and the district in turn assigns children to schools via a centralized mechanism. 
School choice researchers employ \emph{discrete choice} models, statistical models of choices made from slates of discrete options, to describe the preference-generation process by breaking a ranking into a sequence of choices from dwindling choice sets.

Such models are useful for explanation, indirectly identifying the most influential school characteristics in the decision-making of families, saving time and resources in surveying families. 
They can also be used for forecasting and planning potential changes in the district offerings. 
Finally, these models are also central to evaluating changes in school choice mechanisms themselves, as policymakers propose changes to assignment mechanisms with the hope of improving district outcomes. 
In the latter contexts, these models play a role in simulating preferences and assignments, and/or evaluating the resulting welfare of assignment under the proposed mechanism. 
Put simply, better preference models lead to better school choice analyses, and better analysis lead to better childhood educational outcomes.

The widely-used ranked preference models in this space, including the Plackett--Luce ``exploded logit'' model \cite{Plackett1968,Luce1977}, model the process of constructing a preference ranking as a series of independent discrete choices (conditional multinomial logit (MNL) in the case of Plackett-Luce) 
based on school, program, and household attributes.
While many such models are simple and interpretable,
there is long-standing evidence in the discrete choice literature for ranking behavior that is {\it rank-heterogeneous}, meaning that the sequence of choices are driven by different considerations as individuals work down a preference list \cite{Chapman1982, Hausman1987, Fok2012}.
The criteria an agent uses for selecting top-ranked alternatives may differ from those at lower ranks, either due to true preference shifts or behavioral mechanisms such as decision fatigue. 

In this work, we present and evaluate two strategies for incorporating rank-heterogeneity in choice models for school choice. One strategy achieves heterogeneity through a sequential dependence using \emph{context effects}, while the other relies on regularized model stratification.

\subsubsection*{Rank-heterogeneity via context effects.} \emph{Context effects} describe the influence of a particular decision context, including the available or previously-chosen options, on a individual's relative preferences between alternatives. We adapt a previous model of context effects, the context-dependent random utility model (CDM) \cite{Seshadri2019}, to the school ranking setting. The CDM has been used to study ranked preferences \cite{Seshadri2020} by decomposing the ranking process as a series of choices in the context of the dwindling set of items yet to be chosen. We consider a variation of the CDM more natural to the school choice setting: modeling the ranking instead as a series of choices in the context of the \emph{already chosen} items.
Surprisingly, we show that the two modeling approaches (respectively, forward-dependence and backwards-dependence) are equivalent, and opt to use the latter variation when interpreting our results.

\subsubsection*{Rank-heterogeneity via model stratification.} 
An alternative approach to inducing rank-heterogeneity is stratifying the modeling problem by rank position. Simply learning a series of independent models for each rank position, however, can split the data too finely and result in poor generalization. To avoid this pitfall, we apply Laplacian regularization \cite{Tuck2021} to the independent models, with carefully tuned regularization graphs that bring models of adjacent choices close together.

Incorporating context effects and model stratification are not mutually exclusive, and we also evaluate the combination of both approaches in our analysis. Moreover, we perform a series of ablation studies to demonstrate the independent contributions of each approach.
We evaluate these new tools by modeling the preferences for the San Francisco Unified School District (SFUSD) kindergarten programs during the 2017-18 and 2018-19 assignment years. We find that the first strategy (context effects) dramatically lowers out-of-sample negative log likelihood, particularly on down-rank choices, when compared to rank-homogeneous models.
The second strategy (model stratification) delivers more accurate prediction in top choices than a rank-homogeneous model---essentially, by modelling them separately---but otherwise does not appear to produce any significant improvements over the non-stratified baseline. 
Furthermore, we evaluate the performance of our context effect model against a nested MNL model and demonstrate sizable advantages in the school choice setting.

\subsubsection*{Outline.} 
Section~\ref{sec:prelims} introduces notation and definitions used throughout the work.
Section~\ref{sec:sfusd} explains the SFUSD assignment system, its inputs and outputs, and summarizes the data we use for training and evaluation. 
In Section~\ref{sec:models}, we describe the choice models studied in this work, presenting the backwards-dependent context-dependent model (CDM) and the stratified approach with Laplacian regularization. 
Section~\ref{sec:selection} addresses identifiability of the models and details our model optimization framework.
In Section~\ref{sec:results}, we present and discuss the performance of our models; Section~\ref{sec:conclusion} concludes.

\subsection{Related Work}
\label{sec:rel}
The present work closely relates to various prior works that develop or apply preference models in school choice.
Laverde \cite{Laverde2022} uses an MNL choice model to simulate counterfactual assignments in Boston in 2010--2013, quantifying the role of distance and unequal access on stated preferences. 
Agarwal \& Somaini~\cite{Agarwal2018} develop a procedure for estimating an MNL model in the presence of strategic reporting. 
Abdulkadiro\u{g}lu et al.~\cite{Pathak2020} use MNL models to find links between preferences, school effectiveness and peer quality in New York City in 2003--2013.
For an in-depth review of prior applications of preference models in school choice, see Agarwal \& Somaini~\cite{Agarwal2020}.

Meanwhile, many works have studied the relative suitability of different choice models in school choice, evaluating accuracy and prediction errors of preference models. For example, Pathak \& Shi \cite{Pathak2017} examine out-of-sample estimates for three models after a large-scale policy change in Boston. 
They develop several model evaluation metrics, and we adapt one to our work. 
Calsamiglia et al.~\cite{Calsamiglia2020} similarly estimate a full choice system and evaluate it out-of-sample using administrative data from 2006 and 2007 school years in Barcelona.

Several prior efforts aim to understand preference heterogeneity between various demographic groups.
For example, Laverde \cite{Laverde2022} estimates MNL models for White, Black, and Hispanic families by including indicator variables for these features in the chosen MNL utility. 
Hastings et al. \cite{Hastings2008} apply mixed-logit models \cite{McFadden2000} to data from Charlotte-Mecklenburg, North Carolina, learning separate model coefficients by race and SES status.
In contrast to these examples of heterogeneity \emph{between} groups, the present work focuses instead on preference heterogeneity \emph{within} participants as they assemble their rankings.

Our idea is inspired by prior works in psychology, economics and marketing research, all of which cite inconsistent agent behavior in the assembly of rankings.
Under the observation that individuals are generally more careful in reporting their top choices than lower ranked ones, Hausman \& Ruud \cite{Hausman1987} model structured rank-heterogeniety through a common choice model with increasing variance as choosers proceed down the ranks,
Chapman \& Staelin \cite{Chapman1982} drop ranked alternatives after a threshold, and 
Allison \& Christakis \cite{Allison1994} interact model covariates with indicators for early (top-4) or late (5+) rank choices.
Our work extends this last idea by fully stratifying models by rank position of choice, interacting all model parameters with indicators for the first $k$ ranks. 
More on our stratification (and regularization) framework in Section~\ref{sec:models}.

Finally, our work applies recent advances from the discrete choice and preference learning literatures to the school choice domain. 
The MNL model satisfies the axiom of independence of irrelevant alternatives (IIA), that the relative probability of selecting any item $j$ over another item $k$ from choice set $S$ is independent of the other items in $S$. However, this axiom is highly restrictive and often not representative of the true choice process \cite{Tversky1969, Tversky1993}. 
We adopt strategies for going beyond the independence of irrelevant alternatives (IIA) assumption from Seshadri et al.~\cite{Seshadri2019}, in turn adapted from Batsell \& Polking \cite{Batsell1985}, extending that framework from a previously-studied forward-dependent model of ranking \cite{Seshadri2020} to backwards-dependent ranking. 
Other recent work extending the CDM include studies of salient features \cite{bower2020preference} and feature-based context effects \cite{tomlinson2021learning}; we leave the evaluation of such model extensions as future work.
Further, we benchmark the performance of our approaches against the nested MNL model \cite{mcfadden1978modelling}, which also goes beyond the restrictive IIA assumption, in Section~\ref{sec:nll}.

\section{Choice Preliminaries}
\label{sec:prelims}

We begin by introducing our notation for viewing school choice through the lens of discrete choice.
For a specific school year, let $\mathcal{U} \coloneqq [m] = \{ 1,...,m \}$ denote the universe set of all offerings, or \emph{alternatives}, in the district, labeled 1 through $m$, and
let $n$ be the number of students seeking assignment in the choice system.
Throughout this work, we use ``household'' and ``student'' interchangeably to represent the decision-maker, as enrollment pertains to the student but rankings are often submitted by caretakers.
Further, let $\text{PO}(\mathcal{U})$ denote the set of all partial orders on the alternatives in $\mathcal{U}$.
A preference list $R_i\in \text{PO}(\mathcal{U})$ is household $i$'s partial ranking of the alternatives in $\mathcal{U}$, and we denote by $k_i\leq m$ the length of that ranking. 
The vector of observable covariates on student $i$ and offering $j\in\mathcal{U}$ are given by $x_{ij}$, containing demographic, socioeconomic, geographic, and performance-related information on the pair.
Then, a \emph{school choice dataset}, $(\mathcal{D}, X)$, is defined as the collection of all participating-household's partial rankings submitted to the district, $\mathcal{D}=\{R_1, ..., R_{n}\}$, and observed student-program covariates, $X\in\mathbb{R}^{n\times m \times d}$, where $x_{ij}$ is a length-$d$ vector of attributes pertaining to student $i$ and alternative $j$.

To learn a model of rank data, researchers typically transform rankings to choices and then apply discrete choice models such as the MNL, resulting in what is known (equivalently) as the \emph{rank-ordered logit} \cite{Hausman1987}, \emph{exploded-logit} \cite{Punj1983, Chapman1982}, or Plackett-Luce \cite{Plackett1968,Luce1977} model for rankings, which we present in Section~\ref{sec:models}.
The generality of converting rankings to choice is non-obvious, but the most powerful and widespread transformation is motivated by the theory of L-decomposable ranking distributions \cite{Critchlow1991,Luce1959} (L as in Left). A ranking distribution is said to be L-decomposable if the probability of observing ranking $R=(r_1,...,r_k)$ can be decomposed into probabilities of choices from dwindling choice sets, from most to least preferred:
\[
P(R) = P(r_1|\{r_1,...,r_k\})P(r_2|\{r_2,...,r_k\})...P(r_{k-1}|\{r_{k-1}, r_k\}).
\]
This unraveling-from-the-left decomposition is sometimes also referred to as \emph{repeated selection} \cite{Seshadri2020}. In the present work, we apply repeated selection to ranking data throughout, simplifying the name of the ranking model to just the enlisted choice model employed after unraveling.

Encoding the unraveled choices as (agent, choice, choice set) triples, the rank data $\mathcal{D}$ then becomes a choice dataset, $D$:
\begin{equation}
D = 
\bigcup_{R_i \in \mathcal{D}}
\bigcup_{j \in [k_i]} 
\left (
i,
r_{ij},
S_{ij}
\right ) 
\label{eq:repeated}
\end{equation}
where $r_{ij}$ is represents the $j$-{th} selection by agent $i$ on ranking $R_i$, and $S_{ij}\subseteq \mathcal{U}$ is the slate of available alternatives, or choice set, when choosing position $j$ of ranking $R_i$. 
The size of the resulting dataset is $|D| = \sum_{i\in[n]}k_i$.

To concretely illustrate the decomposition at the level of a data point, given a universe of alternatives $\mathcal{U}=\{a,b,c,d\}$, consider a dataset made up of one ranking, by agent $1$, $\mathcal{D}=\{R_1\}$, where $R_1=(b,d,c)$.
Following Eq.~\eqref{eq:repeated}, the choice dataset becomes
\begin{align*}
    D&=\{(1,r_{11},S_{11}),(1,r_{12},S_{12}),(1,r_{13},S_{13})\}\\
    &=\{(1,b,\{a,b,c,d\}), (1,d,\{a,c,d\}), (1,c,\{a,c\})\}.
\end{align*}

\section{San Francisco School Choice}
\label{sec:sfusd}
In this section, we present the assignment process implemented within the San Francisco Unified School District (SFUSD) from 2014 to the present. SFUSD is made up of 130 schools with 150+ unique program offerings.
Students enroll in these programs via an annual assignment lottery where families submit ranked preferences over available offerings to the district, and the district tries to honor family choices while satisfying capacity constraints. 
The algorithm performing this constrained assignment is the student-proposing deferred acceptance algorithm \cite{Gale1962}.
Participation may occur across all grade levels, but kindergarten enrollment is by far the largest participating group each year, making up over a third of all annual participants.
As such, and following suit with many other studies of school choice, we focus solely on kindergarten assignment.

In the face of overly-demanded program seats, the district uses the following priority hierarchy to make assignments:
\begin{enumerate}
    \item \textbf{Sibling}: Highest priority. Given to younger siblings of students enrolled at the school.
    \item \textbf{PreK/TK}: Given to students who (1) live in the attendance area of the school (if applicable), and (2) are enrolled in a PreK or TK program at the school itself or in the attendance area of the school (if applicable).
    \item \textbf{Test score area (``CTIP1'')}: Given to students living in neighborhoods with low average test scores. Grants priority across the district, not just to one program or school.
    \item \textbf{Attendance area (AA)}: Given to students living in the attendance area of the school. 
   \item \textbf{No priority}: The absence of any of the above priorities.
\end{enumerate}
For each program, a student is considered in the highest priority category for which they qualify. 
Within each priority tier, ties are broken by next highest tier if applicable, or by a random number, $v_{ij}$, drawn uniformly at random for each student-school pair\footnote{This lottery design is known as the \emph{multiple tie-breaking rule} (MTB) as students receive multiple values, one for each ranked school. By contrast, the single tie-breaking rule (STB) assigns a single lottery value to each student, used across desired schools. 
For more on the analysis of tie-breaking rules, see \cite{Abdulkadiroğlu2009,Haan2015,Ashlagi2019,Ashlagi2020}.}.

Once all submitted preference lists have been exhausted by the matching algorithm, there may be students left without any assignment, for which the district administratively assigns these students to a program not on their list.
In this work, as we are solely interested in modeling the preferences submitted by families in the first stage, such assignments fall outside the scope of our analysis.

\subsection{Dataset}
To understand families stated preferences, we study data from both the 2017--18 and 2018--19 school years, principally training models on the 2017--18 data and evaluating out-of-sample on the 2018--19 data. We opted for this train--test split, following other work in school choice \cite{Calsamiglia2020, Pathak2017}, to prevent data leakage. This split also mimics real use cases of school choice preference models, where models are used to simulate future years' outcomes. 

Within each year, we collapse all three rounds of stated preference elicitation (instead of focusing only on the first and largest round), to better capture the full district demographics. 
We exclude programs that are newly offered the 18-19 school year, dropping 804 households from the test dataset, as our model's fixed effects do not extrapolate to never-before-seen alternatives. Handling these out-of-distribution preferences is a known limitation of the MNL class of models and an area of future work.
Summary statistics for each school year's data are found in Table~\ref{tab:data}.

\begin{table}[t]
    \centering
    \caption{Summary statistics of SFUSD dataset, by school year.}
    \begin{tabular}{l|c|c}
                                            & \multicolumn{2}{c}{School year}\\\cline{2-3}
                                            & 2017-18   & 2018-19 \\\hline
        No. participating households, $n$   & 5,115     & 4,329\\
        Total offerings, $m$                & 154       & 148\\
        No. unique schools, $n_s$           & 72        & 72 \\
        No. unique program types, $n_p$     & 22        & 19 \\
        Avg. length of ranking, $\bar{k}_i$ & 9.95      & 7.05 \\
        Size of choice dataset, $\sum_i k_i$& 49,882    & 29,810 \\
        Percent students CTIP1              & 16.7\%    & 18.7\% 
    \end{tabular}
    \label{tab:data}
\end{table}

The student-program covariates $X$ used in this work were selected by domain experts at SFUSD:
\begin{itemize}
    \item Distance: scalar, in miles,
    \item Square-root distance: scalar, in sqrt. miles,
    \item Square-root distance $\times$ CTIP1: scalar, in sqrt. miles,
    \item Within 0.5 miles: indicator,
    \item Bus route: indicator for whether the district has bus routes between student ZIP code and school ZIP code,
    \item Sibling match: indicator for whether the student has one or more sibling(s) already enrolled at the school (not necessarily same program),
    \item Language match: indicator for whether a language program is in a student's (non-English) home language,
    \item Attendance area school: indicator for whether the student lives in the attendance area of the school,
    \item PreK/TK continuation: indicator for whether or not student is enrolled in an SFUSD Pre-K or transitional kindergarten in the same attendance area as or within the school.
\end{itemize}
We consider the following school-specific features as well, modeled as interacting with the CTIP1-status of the student.
\begin{itemize}
    \item Average color: state-defined metric quantifying school's absolute performance and improvement in English/language arts, math, chronic absenteeism, and suspension rates. Ordinal color code in each category, encoded as 1-5, and averaged (higher is better),
    \item Fraction reduced lunch: fraction of the school's population that qualifies for free or reduced-price lunch by the district,
    \item Before/after school programs: indicator for whether or not school offers before- or after-school programs.
\end{itemize}
We acknowledge that additional attention to feature engineering can likely yield measurable performance improvements, but we consider the above features adequate and realistic for our purposes, namely evaluating the value of modeling rank-heterogeneity.

\section{Choice models for School Choice}
\label{sec:models}

A choice model models probability distributions over subsets of a collection. More formally, let $\mathcal{S} = \{S : S\subseteq\mathcal{U}, |S|\geq2\}$ denote the set of all subsets of size at least two of a collection $\mathcal{U}$. Let $P(j|i,S)$ describe, for each agent $i\in[n]$ and each set $S\in \mathcal{S}$, the probability of agent $i$ selecting item $j$ from set $S$.
Recall from Section~\ref{sec:prelims} that in the SFUSD school choice mechanism, $n$ households submit partial rankings $R_1, R_2, ..., R_{n}$ with student-program covariates $X$.
Each partial ranking $R_i$ is decomposed into choices per Equation~\eqref{eq:repeated}, obtaining a dataset $D$ of choices.

We begin by considering a random utility model (RUM) of choice.
The utility to agent $i$ of alternative $j$ in choice set $S$ is given by $$U(j|i,S) = V(j|i,S) +\epsilon_{ij},$$ decomposed into a part labeled $V$ that is known by the researcher up to some parameters, the \emph{representative utility}, and an unknown part $\epsilon$ that is treated as random \cite{Train09}. 
Under the assumption of independent Gumbel noise $\epsilon$, agent $i$'s probability of choosing alternative $j$ from choice set $S$ is given in closed form by
\begin{equation}
P(j|i,S) = \frac{e^{V(j|i,S)}}{\sum_{k\in S} e^{V(k|i,S)}},
\label{eq:prob}
\end{equation}
deriving the most ubiquitous RUM---especially in school choice---the conditional multinomial logit (MNL) \cite{Luce1959}. 

Taking the noise instead to be \emph{jointly} Gumbel distributed with correlation yields variations on a mixed MNL~\cite{McFadden2000} or nested MNL~\cite{Train09} model, the latter featuring correlations across pre-specified clusters of alternatives. We benchmark our performance against a nested MNL model in Section~\ref{sec:results}. Mixed MNL models have performed comparably to ordinary MNL in several prior school choice studies~\cite{Hastings2008, Pathak2014}, so we do not benchmark against it in this work.

Under the MNL model, the task of the researcher is to define a representative utility function, typically a parametric model, denoted $V_\theta$.
We select our model from the chosen model class using regularized maximum likelihood, selecting parameters $\theta$ to minimize
\begin{equation}
F(D;\theta) = \ell(D; \theta) + r(\theta),
\label{eq:objective}
\end{equation}
where $\ell(D;\theta)$ is the negative log-likelihood (NLL) loss, 
\begin{align*}
    \ell(D;\theta) &= -\frac{1}{|D|}\sum_{(i, r_{ij}, S_{ij})\in D}\log\left(P_\theta(r_{ij}|i,S_{ij})\right),
\end{align*}
$r(\theta) = \lambda||\theta||_2$ is the $\ell_2$ penalty, and $\lambda$ is the regularization gain. 

\subsection{Basic utilities}
\label{sec:base}
At this point, our task is to define the representative utility, $V$. 
Assigning inherent utilities to each alternative,
\begin{equation}
    V_\theta(j|i,S) = \delta_j,
    \label{eq:pl}
\end{equation}
reduces the model to the basic Plackett-Luce model \cite{Luce1959}. Here, $\delta\in\mathbb{R}^{\tilde m}$ where $\tilde{m} < m$ is defined as the number of unique schools plus the number of unique program-types offered in the district\footnote{Each alternative in the choice universe $j\in\mathcal{U}$ has an associated school, $s(j)$, and program type, $p(j)$. Example program types are general education, special education, and language program offerings. As such, our fixed effect $\delta_j$ is actually shorthand for $\delta_{s(j)}+\delta_{p(j)}$, reducing degrees of freedom while allowing our models to better generalize to new offerings at existing schools. We refer the interested reader to our code for the exact implementation.}, $\tilde{m} = n_s+n_p$. See Table~\ref{tab:data} for district summary statistics.
We will refer to this model as the \textbf{fixed-effect MNL}.

Adding user-alternative specific covariates yields a \textbf{linear MNL},
\begin{equation}
    V_\theta(j|i,S) = \delta_j + \beta^Tx_{ij},
    \label{eq:lin}
\end{equation} 
the most common utility structure in the school choice literature. 
Note that covariates contained in $x_{ij}$, detailed in the previous section, are indexed both by student and alternative; school- or program-specific features (only indexed by $j$) are absorbed into the fixed effects $\delta_j$, and student-specific features (only indexed by $i$) cancel out in the expression of the MNL choice probability, Eq.~\eqref{eq:prob}.

As as auxiliary benchmark, we implement a \textbf{nested MNL} model, using the same expression of representative utility as the linear MNL in Eq.~\eqref{eq:lin}, for benchmarking in Section~\ref{sec:results}. 
In the nested MNL, alternatives are explicitly assigned to one of $K$ nests (non-overlapping subsets of $\mathcal{U}$), and choice probabilities are defined to be correlated within nests. 
See Appendix~\ref{sec:nested} for full discussion and presentation of the nested MNL choice probabilities. 
The nests we implement in this work are ‘Chinese Language’, ‘Filipino Language’, ‘General Education’, ‘Japanese Language’, ‘Korean Language’, ‘Spanish Language’, and ‘Special Education’ offerings. Each nest is associated with a number of unique program offerings ---see Table~\ref{tab:data} for more details.

The fixed-effect and linear MNL models presented in this section satisfy IIA (see Section~\ref{sec:rel}). As a result, they are \textit{rank-homogeneous}, relying on a constant representative utility $V(j|i,S;\theta)$ for each alternative regardless of when the choice is being made in the ranking process. The contextual choice model that follows does not satisfy IIA and thus leads to rank-heterogeneous choice distributions.

\subsection{Context effects}
The \textbf{context-dependent model} (CDM) \cite{Seshadri2020} is our first strategy for incorporating rank-heterogeneity into the traditional MNL models above.
The CDM relaxes the strict IIA assumption, initially crafted to model ``choice set effects''~\cite{trueblood2013not} whereby the slate of alternatives under consideration impacts the choice probabilities of the agent. 
We modify this modeling framework to suit the school choice sequential ranking problem.
Specifically, under the standard CDM, each choice $j$ from choice set $S$ occurs within the context of the choice set $S$. For our purposes, we generalize this framework to consider the choices as occurring within the context of a generic and possibly different \emph{context set} of alternatives, $A$. 

The representative utility of this generalized CDM models context effects as a linear dependence between items, interpretable as ``push'' and ``pull'' factors, with items in $A$ pushing and pulling on each alternative in the choice set $S$,
\[
V_\theta(j|i,A,S) = \delta_j + \beta^Tx_{ij} + \frac{1}{|A|}\sum_{k\in A}u_{jk}, \forall j \in S.
\]
When $A=S \setminus j$ we recover the standard CDM. 
The parameters $u_{ij}$ are defined for all $i,j\in\mathcal{U}$ where $i\neq j$. 
The generalized CDM has the same parameter complexity as standard CDM, requiring $m(m-1)$ parameters beyond the linear model, arranged in a matrix-like structure $U\in\mathbb{R}^{m\times m}$, with undefined diagonal. 
To reduce the parametric complexity of the model, $U$ can be factorized as the product of two low-rank matrices, $U=TC^T$ with $T,C\in\mathbb{R}^{m\times r}$ serving as \emph{target} and \emph{context} embeddings, respectively, analogous to word2vec-type methods \cite{mikolov2013}.
The low-rank CDM representative utility is then written as
\begin{equation}
V_\theta(j|i,A,S) = \delta_j + \beta^Tx_{ij} + \frac{1}{|A|}\sum_{k\in A}t_j^Tc_k.
\label{eq:cdm}
\end{equation}
We proceed with the factorized form of the CDM in this work. 
The low-rank CDM introduces a hyperparameter, in the form of the embedding dimension $r$; see Section~\ref{sec:selection} for a discussion of hyperparameter tuning.

To accompany this change in model, the structure of the data described in Eq.~\eqref{eq:repeated} must be generalized to include a generic context set for each choice, resulting in the following choice dataset
\begin{equation}
D = 
\bigcup_{R_i \in \mathcal{D}}
\bigcup_{j \in [k_i]} 
\left (
i,
r_{ij},
A_{ij},
S_{ij}
\right ),
\label{eq:cdm_repeated}
\end{equation}
where $A_{ij}$ is the context set when agent $i$ chose item $j$.
Table~\ref{tab:models} summarizes the representative utilities of the three models---the fixed-effect MNL, linear MNL, and CDM---and their parameters.

\subsubsection*{Forward vs. backward-dependence}
In the original formulation of the CDM for rankings \cite{Seshadri2020}, the context set was assumed to be the choice set itself, $A=S\setminus j$, a formulation we refer to as the \emph{forward-dependent} contextual ranking model. 

Considering the generalized CDM above, we consider instead a model where the context set $A$ is the set of \emph{already-chosen} alternatives. Equivalently, let $A=\mathcal{U}\setminus S$, the complement of the current choice set. We introduce this model as the \emph{backward-dependent} contextual ranking model. Rather than modeling context effects between alternatives in the choice set, it measures how well each alternative fits with the choices already made.
This conceptual shift is better suited to the psychology of the school choice selection process than the former framing, and yields a more interpretable model in ranking settings where choice sets are large, such as in school choice.

Considering these two different approaches to modeling rankings as a sequence of contextual choices, it seems as though these formulations constitute different model classes. However, we find that the forwards- and backwards-dependent CDM ranking model classes are in fact equivalent, and provide a bijection between the spaces of parameters for both the unfactorized and factorized models. 
See Appendix~\ref{sec:equiv} for proofs of Theorems~\ref{thm:full} and \ref{thm:low}.

\begin{theorem}
Let $\theta_F=\{\delta^F, \beta^F, U^F\}$ denote model parameters of the unfactorized forward-dependent CDM ranking model, and $\theta_B$ denote those of the backward-dependent model. 
The forward- and backward-dependent parameters are equivalent under the bijection $\theta_B = f(\theta_F)$, where
\[
f(\theta) = \left\{\Bigl\{\delta_i+\sum_{j\in\mathcal{U}\setminus i} u_{ij},~\forall i\Bigr\}, \beta, -U\right\}.
\]
The inverse map is the map itself, $f^{-1} = f$.
\label{thm:full}
\end{theorem}
\begin{theorem}
Let $\theta_F=\{\delta^F, \beta^F, T^F, C^F\}$ denote model parameters of the low-rank forward-dependent CDM ranking model, and $\theta_B$ denote those of the low-rank backward-dependent model. These model parameters are equivalent under the bijection $\theta_B = g(\theta_F)$, where
\[
g(\theta) = \left\{\Bigl\{\delta_i+t_{i}^T\sum_{j\in\mathcal{U}\setminus i} c_j,~\forall i\Bigr\}, \beta, T, -C\right\}.
\]
The inverse map is the map itself, $g^{-1} = g$.
\label{thm:low}
\end{theorem}

In Section~\ref{sec:results}, as a supplemental analysis, we consider results for truncated top-$k$-dependent context sets, $A_{ij}=\{r_{i1}, ..., r_{il}\}$ for $l=\min(k, j)$, to evaluate whether a more limited dependence (and thus, simpler model) performs as well as full backward-dependence. We find that even the truncated top-1 CDM---with knowledge only of the agent's first choice---makes considerable gains over the linear MNL model, but the full (top-$m$) backwards-dependent CDM exhibits the best performance.

\begin{table}[t]
    \centering
    \caption{Summary of models and the number of degrees of freedom, $N_p$. Here $m$ denotes the number of alternatives (school and program pairs) offered by the district, $\tilde{m} = n_s + n_p$ denotes the total number of unique schools and program types, $d$ is the length of $x_{ij}$, and $r$ is the embedding dimension of the low-rank CDM. 
    }
    \begin{tabular}{r|c|c|c}
        Model & $V_\theta(j|i,A,S)$ & $\theta$ & $N_p$ \\\hline
        Fixed & $\delta_j$ & $\{\delta\}$ & $\tilde{m}$\\
        Linear & $\delta_j + \beta^Tx_{ij}$ & $\{\delta, \beta\}$ & $\tilde{m}+d$\\
        CDM & $\delta_j + \beta^Tx_{ij} + \frac{1}{|A|}\sum_{k\in A}t_j^Tc_k$ & $\{\delta, \beta, T, C\}$ & $\tilde{m}+d+2rm$\\
    \end{tabular}
    \label{tab:models}
\end{table}

\subsection{Stratifying across ranks}
It has been generically noted \cite{Chapman1982, Hausman1987, Fok2012} that the criteria individuals use for selecting top-ranked alternatives differs from those for lower-ranked alternatives, either due to decision fatigue or a true preference shift. 
As such, we consider the possibility of within-agent preference shift by stratifying the choice model by rank, and learning independent choice models at each rank position.
A possible concern with this approach is that we end up with considerably less data for each model in this stratification, compared to estimating a single common model. To address this concern, we encourage models at neighboring ranks to be close to one another via Laplacian regularization, resulting in a \textbf{Laplacian-regularized stratified model} \cite{Tuck2021}. The methods of Laplacian-regularized stratification are closely related to popular methods for smoothing ($\ell_2$) and trend filtering ($\ell_1$) in temporal \cite{kim2009ell1} and general graphical \cite{smola2003kernels,wang2015trend} domains, where the underlying idea of parameter fusion dates back to at least the work of Land and Friedman \cite{land1997variable,tibshirani2005sparsity}.

The stratification builds upon a base choice model---in this work, one of the three models summarized in Table~\ref{tab:models}. 
Taking the number of strata to be $K$, a stratified choice model is then the composition of $K$ sub-models with parameters $\theta = \{\theta_1,...,\theta_K\}\in \mathbb{R}^{K\times N}$, where $N$ is the number of parameters in the chosen base model.

The $K$ models are regularized towards each other as dictated by an accompanying \textit{regularization graph} \cite{Tuck2021}. 
In our case, rank-based stratification lends itself well to a common ``path graph'' for regularization, where models of adjacent ranks are connected by edges and thus regularized towards each other.
Laplacian regularization here is then defined as:
$$
r_\mathcal{L}(\theta) = \lambda_\mathcal{L} \sum_{i=2}^K||\theta_{i}-\theta_{i-1}||_2^2,
$$
where $\lambda_\mathcal{L}$ is a chosen Laplacian regularization strength, and $r_\mathcal{L}$ is convex in $\theta$. 
Compared to the non-stratified objective in Eq.~\eqref{eq:objective}, the regularized, stratified objective function is the sum of $K$ decoupled model losses (each with a local $\ell_2$ regularization) and the Laplacian regularization term:
\begin{equation}
F(D|\theta) = \sum_{k=1}^K \left [ \ell(D_k;\theta_k) + r(\theta_k) \right ] + r_\mathcal{L}(\theta).
\label{eq:strat_loss}
\end{equation}

Regularized stratified models feature two additional hyperparameters over their base models, the number of strata $K$ and the Laplacian regularization gain $\lambda_\mathcal{L}$; see Section~\ref{sec:selection} for a discussion of hyperparameter tuning.

\section{Model Selection and Optimization}
\label{sec:selection}
We briefly discuss the identifiability of the presented models, alongside details about hyperparameter tuning and optimization.
A model is \emph{identifiable} if no two distinct sets of parameters, $\theta$ and $\theta'$, produce the same probability distributions over all choice sets $S\in\mathcal{S}$.
Identifiability is crucial in settings where decisions are made based on interpreting parameter estimates. If the goal is solely to make decisions based on the resulting distributions only, e.g., from predictions or simulations, identifiability is not strictly necessary.

The traditional MNL family of ranking distributions are non-identifiable due to their shift-invariance.
In this case, strategies for achieving identifiability are to fix one of the parameters, constrain their sum, or to apply regularization and obtain the minimum-norm parameter estimates \cite{Xia2019}.
In our work, we employ the latter strategy for the MNL and all other models, applying non-zero $\ell_2$ regularization, $r(\theta)$, in the objective function and achieving identifiability by obtaining the minimum norm solution. 

The models in this work introduce additional hyperparameters; the low-rank CDM requires the selection of the embedding dimension $r$, and a stratified model is specified by $K$ and $\lambda_{\mathcal{L}}$; the number of strata and amount of Laplacian regularization, respectively.
We tune these hyperparameters via 5-fold cross validation within our training dataset, selecting the values that minimize validation loss. 
Figures illustrating our search over these hyperparameters are found in Appendix~\ref{sec:hyper}, with Table~\ref{tab:hyperparams} summarizing the chosen values.
With hyperparameter values selected and regularization in place, the models are fully specified and we proceed to train our models on the full 2017--18 dataset for testing on the 2018--19 dataset.

\begin{figure*}[t!]
    \centering
    \includegraphics[width=\textwidth]{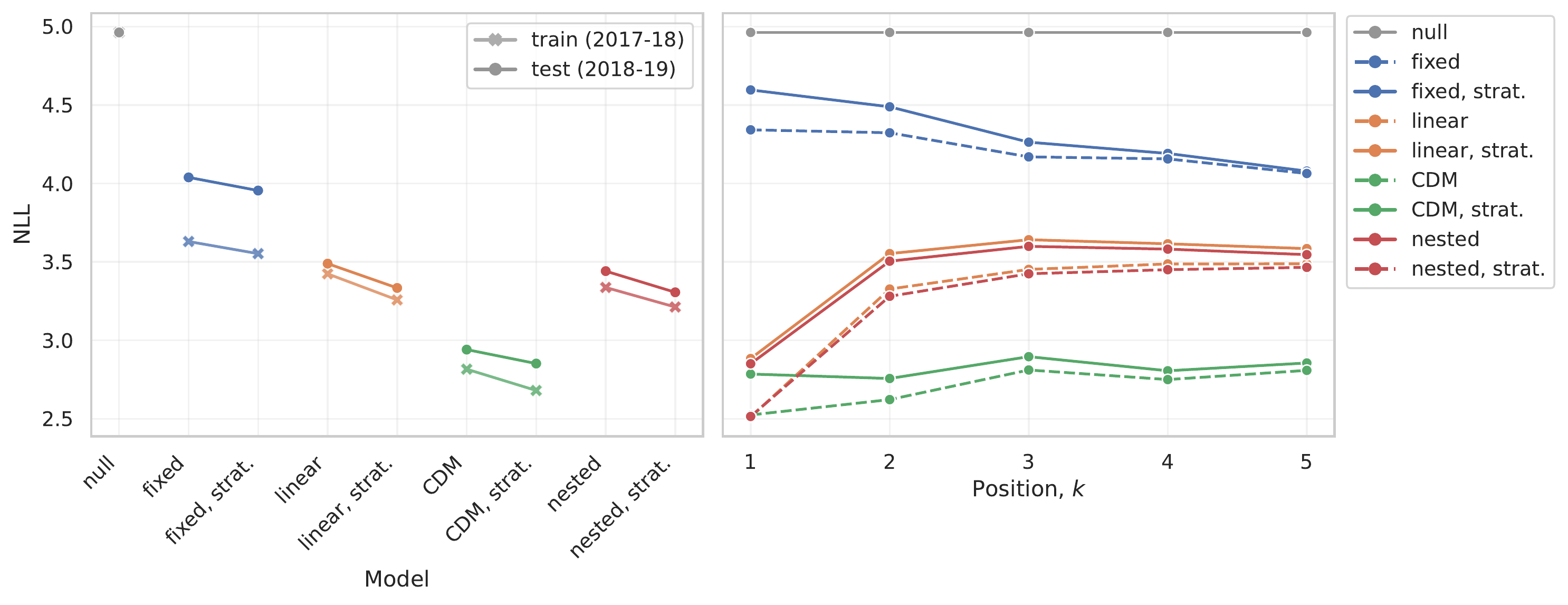}
    \caption{
     Model negative log-likelihoods. Left plot shows overall train and test losses (lower is better) on train (2017-18) and test (2018-19) datasets, respectively. Right plot shows out-of-sample test (2018-19) losses disaggregated by rank. CDM models display lowest overall test losses overall and down rank, and stratified models (linear MNL or CDM) fit top-choices best.}
    \label{fig:overall}
\end{figure*}

We run Adam \cite{Kingma2014}, implemented in PyTorch, with default parameters, $\text{lr} = 0.001$, $\beta = (0.9, 0.999)$, $\epsilon= 1e-8$, adding $\ell_2$ regularization with weight $\lambda= 1e-5$ in accordance with our hyperparameter selection for $\lambda$. 
Model parameters are updated over batches of training data until reaching $\texttt{max\_epoch}=1000$ or convergence, i.e., when the absolute difference in losses is less than $\epsilon=1e-4$. See Appendix~\ref{sec:params} for a discussion on the learned model parameters.

\section{Results}
\label{sec:results}
In this section, we evaluate and examine eight models---non-stratified and stratified versions of the fixed-effect MNL, linear MNL, CDM, and nested MNL models---trained on 2017--18 preference data and evaluated out-of-sample on 2018--19 data.
We observe unique advantages of the context effects modeled by the CDM when benchmarked against the other models, and find that stratifying any model results in strictly (but marginally) better predictions, mostly for top (first) choices.

\subsection{Goodness of fit}
\label{sec:nll}
Figure~\ref{fig:overall} depicts train and test negative log likelihood (NLL) losses on the left, and test losses disaggregated by rank on the right. 
We include a ``null'' model in the plots, representing uniform choices over programs, as a baseline reference point.
We see that the CDM models, stratified or not, result in considerably lower test losses than the fixed-effect, linear, and nested MNL models overall. 
Stratifying provides modest decreases in overall test loss across all models.

On top choices, many families have priority access to one school in the district (e.g., sibling or PreK/TK priorities), and in most cases, rank these schools first. The linear and CDM models incorporate these priorities into the model, and therefore model top choices better than the fixed-effect model. The CDM leverages no additional information in the first choice as the context set is empty (i.e., no choices have been made). 
As such, there is negligible difference between the linear MNL and CDM models at position 1. 
However, after the relatively easy task of predicting top-choices, the CDM is able to leverage the choices made and separates itself from the lower-fidelity models. 
Stratifying yields a lower test loss for top choice across all three models, but quickly loses its advantage at lower ranks, likely due to diminishing training data at those positions (Cf.\ Table~\ref{tab:data}, households rank fewer than 10 programs on average).

\subsubsection*{Truncated top-$k$-dependent CDM}
Recall from Eq.~\eqref{eq:cdm} that the CDM utility differs from the linear MNL model via a sum of pairwise interactions between alternatives and a context set, $A$. 
Throughout this work, the context set is taken to be the set of all previously-chosen alternatives, which has a powerful equivalence in expressivity (Theorems \ref{thm:full} and \ref{thm:low}) to the standard CDM. 
As a robustness check, it is reasonable to ask which prior choices are \emph{most relevant} to the context set. 
We evaluate several variations on the CDM model with the context being defined as the set of $k$ top alternatives. 
Specifically, the utility is given by Eq.~\eqref{eq:cdm} with $A_{ij}=\{r_{i1},...,r_{il}\}$ for agent $i$'s $j$-th choice, where $l=\min(j,k)$. 
In other words, for choices made after position $k$, only the first $k$ choices constitute the context.

\begin{figure}[t]
    \centering
    \includegraphics[width=0.9\columnwidth]{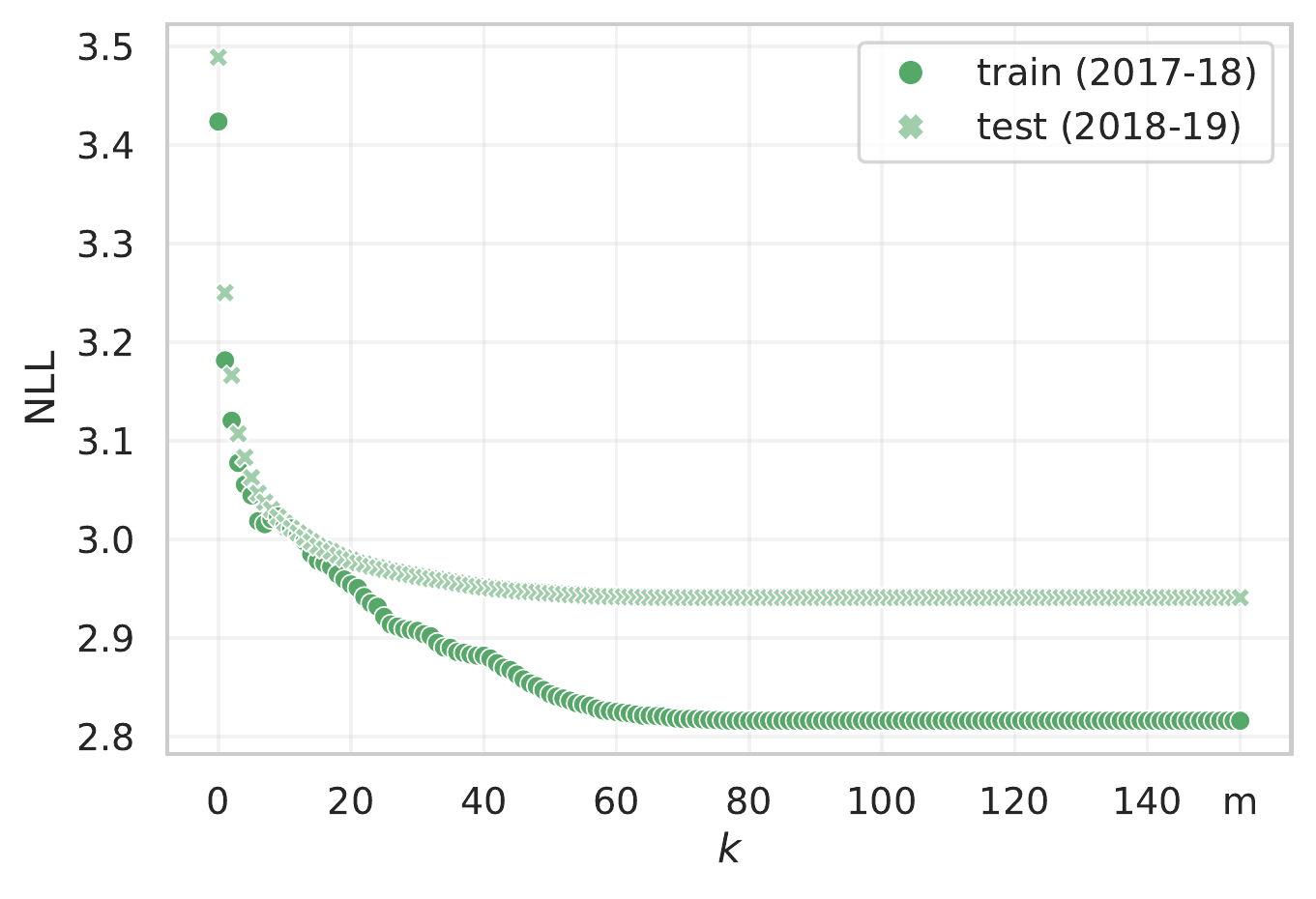}
    \caption{Negative log likelihoods of truncated top-$k$-dependent CDMs where the context set is only the top-$k$ chosen alternatives. Here $k=0$ is equivalent to the linear MNL, and $k=m$, the total number of offered programs, is equivalent to the standard CDM.}
    \label{fig:topk}
\end{figure}

Figure~\ref{fig:topk} presents the losses for these top-$k$-dependent CDM models. 
When $k=0$, the context set is always empty and the model is equivalent to the linear MNL. 
When $k=m$, the number of offered programs, we recover the backwards-dependent CDM considered everywhere else in this work. 
We find that even a minimal context set, e.g., the top-choice only ($k=1$), provides considerable improvement compared to the no-context linear model.
That is, the information of what an agent chose first supplies the model with meaningful signal in making all down-rank predictions. 
That said, letting the context effect be linear in the full set of prior choices ($k=m$) has measurable advantages.

\subsubsection*{Interpreting context effects}
\begin{figure}
    \centering
    \includegraphics[width=0.9\columnwidth]{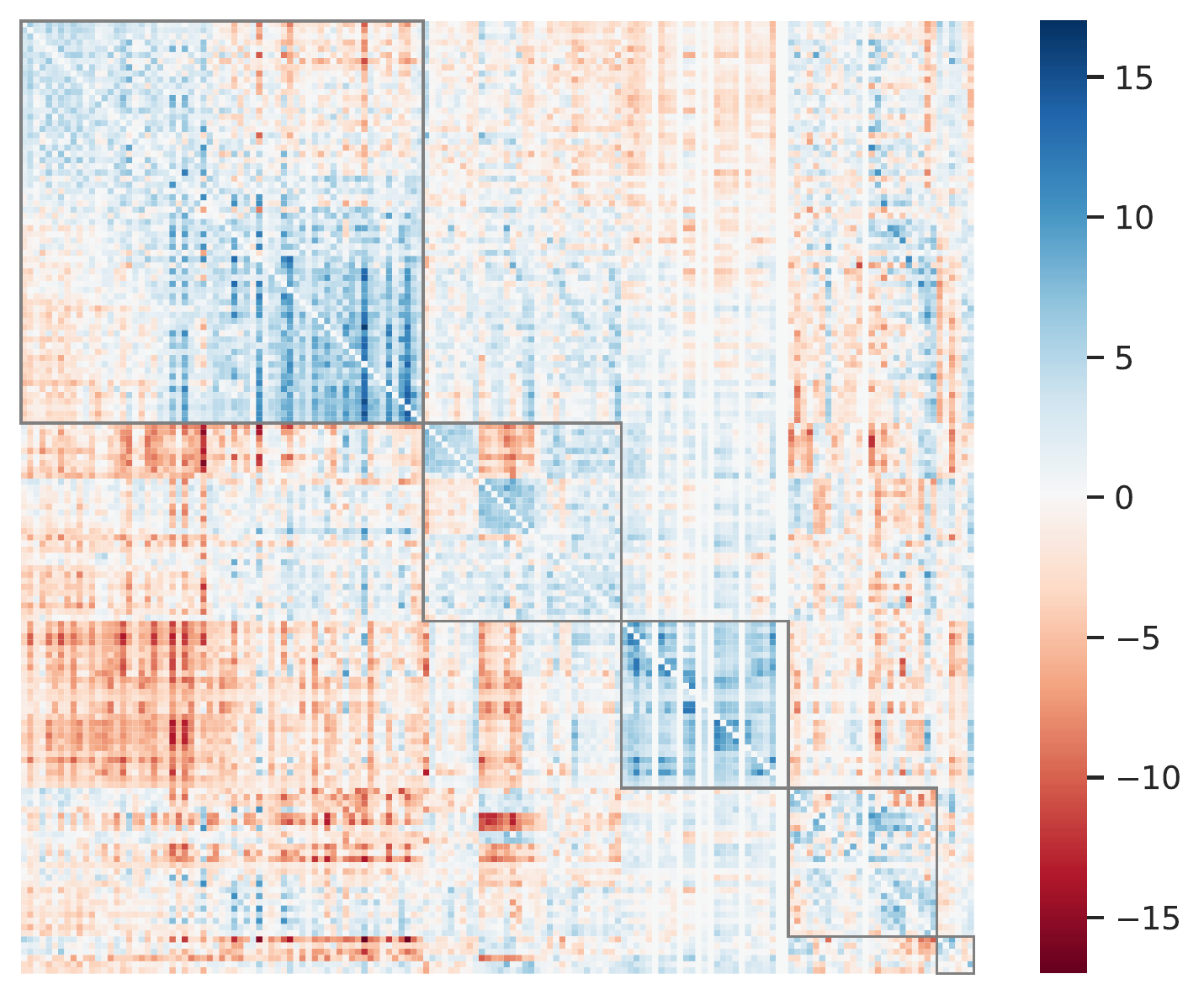}
    \caption{The context effect matrix, $U=TC^T$, of the CDM. Element $u_{ij}$ is the utility boost program $j$ receives from program $i$ being in the context set.
    The block diagonal structure, highlighted with grey outlines, suggests that the CDM primarily (but not only) learns effects between like program types.}
    \label{fig:U_col}
\end{figure}
In Figure~\ref{fig:U_col},
we show the pairwise interactions $U=TC^T$ estimated for the (non-stratified) backwards-dependent CDM. Element $u_{ij} = t_j^T c_i$ is the utility boost that program $j$ receives from chosen program $i$ being in the context set. In the heatmap, programs on the $x$- and $y$-axes were arranged first by program type and then by (descending) popularity within each. From top/left to bottom/right, the program types are General Education (65), Spanish Language (32), Special Education (27), Chinese Language (24), and Miscellaneous Language (6) programs.

We see significant block structure in the matrix, suggesting that the CDM primarily (but not only) uses the context set to learn program type affinities. For example, the third block along the diagonal corresponds to Special Education programs, where we see a strong positive context effect. That is, once a family has ranked a special education program, it becomes much more likely that the family will rank other special education programs. This model behavior is highly intuitive, and is also beyond the behavior of an MNL model or any other model assuming independence. 
Put simply, the CDM's use of context effects enables it to pick up on household signals, from the second choice and onward, that are otherwise not available \emph{a priori} at the household level.

\begin{figure}[t]
    \centering
    \includegraphics[width=0.9\columnwidth]{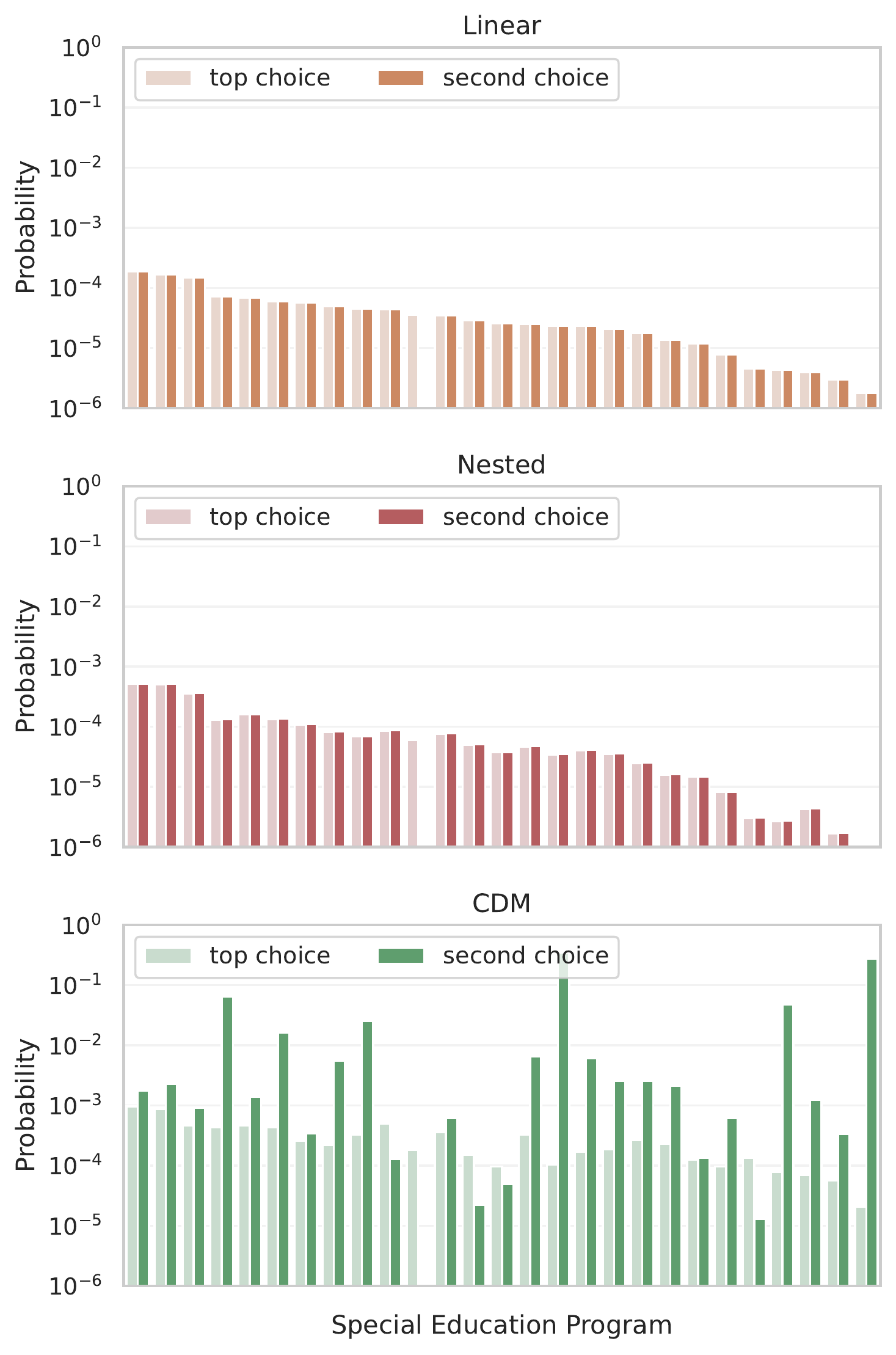}
    \caption{
    Top- and second-choice probabilities of the non-stratified linear, nested, and CDM models over special education programs for a sample household whose first choice was a special education program. The CDM model learns a drastically updated second-choice probability given the context of the top choice.}
    \label{fig:nested_probs_speced}
\end{figure}

The block structure of $U$ may seem to suggest good performance from a nested MNL model, as the latter explicitly clusters similar programs. Instead, in Figure~\ref{fig:overall} we find that the nested MNL shows only marginal gains over the linear MNL model and is not competitive with the CDM. This result sounds surprising, but is fairly intuitive;
in models obeying IIA (such as the fixed-effect and linear MNL models), when an item is removed from the choice set, that item's probability is proportionally redistributed to the remaining alternatives for follow-up choices. The nested model instead allows the removed alternative’s probability to be \emph{non-proportionally} distributed to the remaining items, specifically by favoring the alternatives in its nest (see Appendix~\ref{sec:nested} for details). However, in this setting, the choice universe and nests are relatively large, so the impact of redistributing already-small choice probabilities is marginal.

To illustrate how choice probabilities are redistributed in different models, Figure~\ref{fig:nested_probs_speced} showcases first- and second-choice probabilities by the non-stratified linear MNL, nested MNL, and CDM models over the special education subset for an example household in the district \emph{who first chose a special education program}.
We see that the CDM drastically alters its second-choice distribution, (correctly) boosting the likelihood of this household choosing another special education program, 
while the nested model's top- and second-choice distributions are almost indistinguishable. 
Special education programs are low-probability selections in the data at large, and the nested paradigm can only marginally influence future predictions when one such item is removed from the choice set. 
The CDM has a far greater ability to update its future distributions in the context of rare chosen items.

\subsection{Down-rank prediction accuracy}
\label{sec:accuracy}
Beyond in- and out-of-sample goodness of fit, we now consider the prediction quality of the models on the test dataset.
Specifically, we task the models with making a prediction at rank position $k$, conditional on the first $k-1$ choices made, resulting in an ``accuracy in $k$th Prediction'' evaluation metric.
\begin{figure}
    \centering
    \includegraphics[width=\columnwidth]{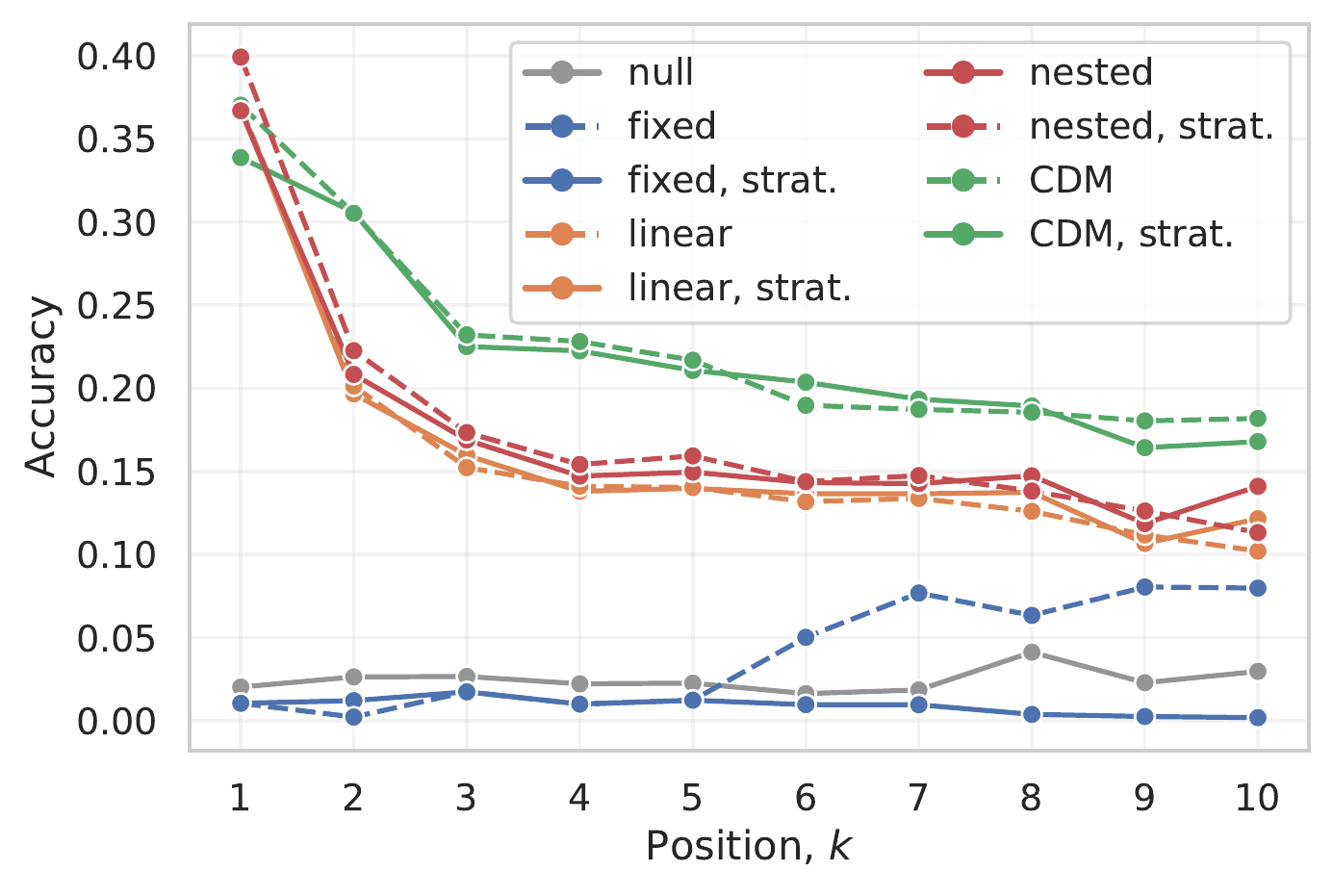}
    \caption{Accuracy in $k$th Prediction. The CDM model makes use of the provided context and generates the most accurate predictions at lower rank positions.}
    \label{fig:completion}
\end{figure}
Recall that $R_i$ denotes the true ranking of household $i$, where $r_{ij}$ is the $j$th item in the ranking for $j\leq k_i$. Let $R_{ik}$ be the set of their true top-$k$ choices, $R_{ik} = \{r_{i1}, r_{i2}, ..., r_{ik}\}$.
Denote by $I_k=\{i : k_i\geq k\}$ the set of households who have ranked at least $k$ alternatives.
Then, given a choice model, denote by $y_{ik}$ the \emph{modal} prediction by the model at position $k$, i.e., the highest probability alternative over remaining programs, in the context of the previous $k-1$ choices, $A_{ik} = R_{i,k-1}$, 
$$
y_{ik}=\arg\max_{j\in S_{ik}}~V_\theta(j|i, A_{ik}, S_{ik}),
$$
where the representative utilities $V(\cdot)$ are defined in Section~\ref{sec:models}.
The metric is then given by
\[
\textbf{Accuracy in $k$th Prediction} = \frac{1}{|I_k|}\sum_{i\in I_k} 1(y_{ik}=r_{ik}).
\]

Figure~\ref{fig:completion} summarizes model performances on this metric.
The CDM models are significantly more accurate in making down-rank predictions when given earlier choices, which is precisely the use-case of the contextual model. 
Stratification leads to improvements in down-rank predictions made by the fixed-effect MNL model, but has limited effect on the linear, nested and CDM models. 
It appears to learn that if a household has not already ranked the most popular programs, they wont be adding them later, as seen in Figure~\ref{fig:strat_logits} of Appendix~\ref{sec:params}. Doing so, it outperforms its non-stratified counterpart beyond position 5.

We can also disaggregate these accuracies by sub-populations of interest, see Appendix~\ref{sec:subpop}. We find that the groups receiving sibling and PreK/TK priorities have top choices that are relatively easy to predict, as their preferences are concentrated on their (typically singular) priority schools. 
All models generally under-perform on CTIP1, Hispanic/Latino, and Black student populations, relative the broader population, for one of two reasons: either the subgroups demonstrate more varied preferences than other subgroups, or the training data was relatively small. Lastly, we see in Figure~\ref{fig:by_subpop_prog} that the CDM specializes in predicting down-rank choices for households with non-mainstream initial preferences.

\section{Conclusion}
\label{sec:conclusion}
In this work, we introduce rank-heterogeneous preference modeling for school choice and present two strategies, discrete choice context effects via a backwards-dependent CDM, and model stratification by rank position. 
Rank-heterogeneous models have the potential to leverage already-chosen alternatives when making down-rank predictions, or to broadly capture evolving household values down a ranking. 
We define and evaluate several metrics, finding that incorporating context terms in the utility dramatically decreases test loss over the linear and nested MNL models, capturing signals not present in covariates alone while also seeing particular improvements in modeling rare choices.
The contextual model also generates more accurate predictions for list-completion tasks. 
Stratifying by rank yields improvements in top-choice accuracy across all models, but otherwise does not result in significant improvements or additional predictive power down-rank. 

While rank-heterogeneous models enable school choice researchers to improve predictions and perform counterfactual analysis, our methods do not come without limitations. For one, the increased parametric complexity of the CDM and regularized stratification strategies raises, albeit mildly, model training times and data requirements relative to the MNL. Recent developments to the CDM \cite{tomlinson2021learning} mitigates this problem by leveraging the model's block structure and learning interactions between program \emph{attributes} rather than the programs themselves. Applying this work to the school choice setting would reduce complexity while uncovering context effects, a promising direction for future work. Another limitation stems from our model failing to generalize to new program offerings with undefined fixed-effects. Here, applying strategies for out-of-distribution prediction---such as establishing a prior on the fixed-effects of new program offerings based on the values for similar offerings---provide further directions for future work. Despite these limitations, we strongly encourage school choice researchers to consider rank-heterogeneous models in their preference modeling tasks for improved down-rank and rare-event prediction.

\xhdr{Reproducibility}
The SFUSD data used in this work is not public, but implementations of all models as well as notebooks used to generate plots in this paper are available at: \url{https://github.com/ameloa/rankingmodels}.
\begin{acks}
We thank the San Francisco Unified School District for providing us access to choice data, specifically thanking SFUSD representatives Joseph Monardo, Jennifer Lebarre, Lauren Koehler, and Reed Levitt for helpful discussions. This work was supported in part by a gift from the Koret Foundation.
\end{acks}

\bibliographystyle{ACM-Reference-Format}
\bibliography{sample-base}

%\newpage

\clearpage
\appendix

\onecolumn

\section{CDM forwards and backwards equivalence}
\newtheorem*{T1}{Theorem~\ref{thm:full}}
\label{sec:equiv}
\begin{T1}
Let $\theta_F=\{\delta^F, \beta^F, U^F\}$ denote model parameters of the unfactorized forward-dependent CDM ranking model, and $\theta_B$ denote those of the backward-dependent model. The forward- and backward-dependent parameters are equivalent under the bijection $f(\theta_F) = \theta_B$, where
\[
f(\theta) = \left\{\Bigl\{\delta_i+\sum_{j\in\mathcal{U}\setminus i} u_{ij},~\forall i\Bigr\}, \beta, -U\right\}.
\]
The inverse map is the map itself: $f^{-1}(\theta_B) = f(\theta_B) = \theta_F$.
\end{T1}

\textbf{Proof}
The latter statement follows immediately from Lemma \ref{lem:full}. Consider the full forwards-dependent CDM: given a set of model parameters $\{\delta^F, \beta^F, U^F\}$ for $\delta^F\in\mathbb{R}^{n_s+n_p}$, $\beta^F\in\mathbb{R}^d$, and $U^F \in \mathbb{R}^{m \times m}$, we have that 
\begin{align*}
    P(j|i,S) = \frac{\exp(\delta_j^F + \beta^{FT}x_{ij}+\sum_{l \in S \setminus{j}} u^F_{jl})}{\sum_{k \in S} \exp(\delta_k^F + \beta^{FT}x_{ik} + \sum_{l \in S\setminus{k}} u^F_{kl})}.
\end{align*}
where $u_{ij}^F$ corresponds to the element in $U^F$ at the row index position corresponding to item $i$ and column index position corresponding to item $j$. The context of relevance here are the other items in the choice set, and that the choice of an item $j$ from a set $S$ is related to how $j$ interacts with that context, in addition to the item fixed-effect $\delta_j^F$ and interactions with agent covariates via the linear term. 

Consider now the full backwards-dependent CDM: given a set of model parameters $\{\delta^B, \beta^B, U^B\}$ for $\delta^B\in\mathbb{R}^{n_s+n_p}$, $\beta^B\in\mathbb{R}^d$, and $U^B \in \mathbb{R}^{m \times m}$, we have that 
\begin{align*}
    P(j|i,S) = \frac{\exp(\delta^B_j + \beta^{BT}x_{ij}+\sum_{l \in \mathcal{U}\setminus S} u_{jl}^B)}{\sum_{k \in S} \exp(\delta_k^B + \beta^{BT}x_{ik} + \sum_{l \in \mathcal{U}\setminus S} u_{kl}^B)}.
\end{align*}
Unlike before, the context of relevance here are items \emph{not} in the choice set--in our case, those that were already chosen--and that the choice of an item $j$ from a set $S$ is related to how $j$ interacts with that context (along with the item fixed-effect and linear interactions, as before).

The forward-dependent CDM may seem like a model class that models different choice probabilities than the backward-dependent model, but the two classes are, in fact, identical. That is, the two model classes model the same collection of choice systems, and have a mapping from one to another. We will show this below. Beginning with the forward-dependent model:
\begin{align*}
    P(j|i,S) &= \frac{\exp(\delta_j^F + \beta^{FT}x_{ij}+\sum_{l \in S \setminus{j}} u^F_{jl})}{\sum_{k \in S} \exp(\delta_k^F + \beta^{FT}x_{ik} + \sum_{l \in S\setminus{k}} u^F_{kl})}.\\
    &= \frac{\exp(\delta_j^F + \beta^{FT}x_{ij}+\sum_{l \in \mathcal{U} \setminus{j}} u^F_{jl}-\sum_{l \in \mathcal{U} \setminus{j}} u^F_{jl}+\sum_{l \in S \setminus{j}} u^F_{jl})}{\sum_{k \in S} \exp(\delta_k^F + \beta^{FT}x_{ik} + \sum_{l \in \mathcal{U}\setminus{k}} u^F_{kl} - \sum_{l \in \mathcal{U}\setminus{k}} u^F_{kl}+ \sum_{l \in S\setminus{k}} u^F_{kl})}.\\
    &= \frac{\exp(\delta_j^F +\sum_{l \in \mathcal{U} \setminus{j}} u^F_{jl} + \beta^{FT}x_{ij}-\sum_{l \in \mathcal{U} \setminus{S}} u^F_{jl})}{\sum_{k \in S} \exp(\delta_k^F + \sum_{l \in \mathcal{U}\setminus{k}} u^F_{kl} + \beta^{FT}x_{ik} - \sum_{l \in \mathcal{U}\setminus{S}} u^F_{kl})}.\\
    &=\frac{\exp(\delta^B_j + \beta^{BT}x_{ij}+\sum_{l \in \mathcal{U}\setminus S} u_{jl}^B)}{\sum_{k \in S} \exp(\delta_k^B + \beta^{BT}x_{ik} + \sum_{l \in \mathcal{U}\setminus S} u_{kl}^B)}.
\end{align*}
where the first statement is the definition of the forward-dependent model, the second adds and subtracts $\sum_{l \in \mathcal{U} \setminus{j}} u^F_{jl}$ from the numerator and $\sum_{l \in \mathcal{U}\setminus{k}} u^F_{kl}$ from the denominator, the third collects and rearranges terms, and the finally line follows by setting 
\begin{align*}
&\delta^B_i := \delta^F_i+\sum_{j\in\mathcal{U}\setminus i} u_{ij}^F, \forall i\\
&\beta^B := \beta^F \\
&U^B := -U^F.
\end{align*}
We observe that the last line is the backwards-dependent CDM, showing that any forwards-dependent CDM can be mapped to a backwards-dependent CDM. Moreover, since the mapping between $\{\delta^F, \beta^F, U^F\}$ and $\{\delta^B, \beta^B, U^B\}$ is $f$, from Lemma \ref{lem:full}, we know that $f^{-1}$ exists (and is $f$), and hence any backwards-dependent CDM can be mapped to a forwards-dependent CDM. This concludes the proof.

\newtheorem*{T2}{Theorem~\ref{thm:low}}
\begin{T2}
Let $\theta_F=\{\delta^F, \beta^F, T^F, C^F\}$ denote model parameters of the low-rank forward-dependent CDM ranking model, and $\theta_B$ denote those of the low-rank backward-dependent model. These model parameters are equivalent under the bijection $g(\theta_F) = \theta_B$, where
\[
g(\theta) = \left\{\Bigl\{\delta_i+t_{i}^T\sum_{j\in\mathcal{U}\setminus i} c_j,~\forall i\Bigr\}, \beta, T, -C\right\}.
\]
The inverse map is the map itself: $g^{-1}(\theta_B) = g(\theta_B) = \theta_F$.
\end{T2}

\textbf{Proof}
The proof follows the same structure as Theorem \ref{thm:full}. The latter statement of the theorem follows immediately from Lemma \ref{lem:low}. We begin with the factorized forwards-dependent CDM: given a set of model parameters $\{\delta^F, \beta^F, T^F, C^F\}$ for $\delta^F\in\mathbb{R}^{n_s+n_p}$, $\beta^F\in\mathbb{R}^d$, and $T^F, C^F \in \mathbb{R}^{m \times r}$, consider the low-rank forwards-dependent CDM model:
\begin{align*}
    P(j|i,S) = \frac{\exp\left(\delta^F_j + \beta^{FT}x_{ij}+t_j^{FT}\left(\sum_{l \in S \setminus{j}} c_l^F\right)\right)}{\sum_{k \in S} \exp\left(\delta^F_k + \beta^{FT}x_{ik} + t_k^{FT}\left(\sum_{l \in S\setminus{k}} c_l^F\right)\right)}.
\end{align*}
We have
\begin{align*}
    P(j|i,S) &= \frac{\exp\left(\delta^F_j + \beta^{FT}x_{ij}+t_j^{FT}\left(\sum_{l \in \mathcal{U} \setminus{j}} c_l^F\right)-t_j^{FT}\left(\sum_{l \in \mathcal{U} \setminus{j}} c_l^F\right)+t_j^{FT}\left(\sum_{l \in S \setminus{j}} c_l^F\right)\right)}{\sum_{k \in S} \exp\left(\delta^F_k + \beta^{FT}x_{ik} +t_k^{FT}\left(\sum_{l \in \mathcal{U}\setminus{k}} c_l^F\right)-t_k^{FT}\left(\sum_{l \in \mathcal{U}\setminus{k}} c_l^F\right)+ t_k^{FT}\left(\sum_{l \in S\setminus{k}} c_l^F\right)\right)}\\
    &= \frac{\exp\left(\delta^F_j +t_j^{FT}\left(\sum_{l \in \mathcal{U} \setminus{j}} c_l^F\right) + \beta^{FT}x_{ij}-t_j^{FT}\left(\sum_{l \in \mathcal{U} \setminus{S}} c_l^F\right)\right)}{\sum_{k \in S} \exp\left(\delta^F_k +t_k^{FT}\left(\sum_{l \in \mathcal{U}\setminus{k}} c_l^F\right) + \beta^{FT}x_{ik} - t_k^{FT}\left(\sum_{l \in \mathcal{U}\setminus{S}} c_l^F\right)\right)}\\
    &= \frac{\exp\left(\delta^B_j + \beta^{BT}x_{ij}+t_j^{BT}\left(\sum_{l \in \mathcal{U} \setminus{S}} c_l^B\right)\right)}{\sum_{k \in S} \exp\left(\delta^B_k + \beta^{BT}x_{ik} - t_k^{BT}\left(\sum_{l \in \mathcal{U}\setminus{S}} c_l^B\right)\right)}.
\end{align*}
where the first statement adds and subtracts $t_j^{FT}\left(\sum_{l \in \mathcal{U} \setminus{j}} c_l^F\right)$ from the numerator and $t_k^{FT}\left(\sum_{l \in \mathcal{U}\setminus{k}} c_l^F\right)$ from the denominator, the second collects and rearranges terms, and the final line follows by setting 
\begin{align*}
&\delta^B_i := \delta_i^F+t_{i}^{FT}\sum_{j\in\mathcal{U}\setminus i} c_j^F,~\forall i \\
&\beta^B := \beta^F \\
&T^B := T^F \\
&C^B := -C^F
\end{align*}

We observe that the last line is the backwards-dependent factorized CDM, showing that any forwards-dependent factorized CDM can be mapped to a backwards-dependent factorized CDM. Moreover, since the mapping between $\{\delta^F, \beta^F, T^F, C^F\}$ and $\{\delta^B, \beta^B, T^B, C^B\}$ is $g$, from Lemma \ref{lem:low}, we know that $g^{-1}$ exists (and is $g$), and hence any backwards-dependent factorized CDM can be mapped to a forwards-dependent factorized CDM. This concludes the proof.
\begin{lemma}
Let $\theta=\{\delta, \beta, U\}$ denote model parameters of the unfactorized forward-dependent CDM ranking model, and let 
$$f(\theta) = \left\{\Bigl\{\delta_i+\sum_{j\in\mathcal{U}\setminus i} u_{ij},~\forall i\Bigr\}, \beta, -U\right\}.$$
The inverse map is the map itself: $f^{-1} = f(\theta)$
\label{lem:full}
\end{lemma}
\textbf{Proof.}
If $f(f(\theta)) = \theta, \forall \theta$, then $f^{-1} = f(\theta)$. We show the former to be true. Let $\theta_F := \{\delta^F, \beta^F, U^F\}$ and let $\theta_B := f(\theta_F) = f(\delta^F, \beta^F, U^F)$. We have,

\begin{align*}
\theta_B=\left\{\Bigl\{\delta^F_i+\sum_{j\in\mathcal{U}\setminus i} u_{ij}^F,~\forall i\Bigr\}, \beta^F, -U^F\right\} = \{\delta^B, \beta^B, U^B\},\\
\end{align*}
where 
\begin{align*}
&\delta^B_i := \delta^F_i+\sum_{j\in\mathcal{U}\setminus i} u_{ij}^F, \forall i\\
&\beta_B := \beta_F \\
&U^B := -U^F
\end{align*}
Now,
\begin{align*}
f(f(\theta_F)) = f(\theta_B)=f(\{\delta^B, \beta^B, U^B\}) &= 
\left\{\Bigl\{\delta^B_i+\sum_{j\in\mathcal{U}\setminus i} u_{ij}^B,~\forall i\Bigr\}, \beta^B, -U^B\right\} \\
&= \left\{\Bigl\{\delta^F_i+\sum_{j\in\mathcal{U}\setminus i} u_{ij}^F+\sum_{j\in\mathcal{U}\setminus i} -u_{ij}^F,~\forall i\Bigr\}, \beta^F, --U^F\right\}.\\
&=\left\{\Bigl\{\delta^F_i,~\forall i\Bigr\}, \beta^F, U^F\right\}.\\
&=\{\delta^F, \beta^F, U^F\} = \theta_F.
\end{align*}
where the first line follows from applying the definition of $f(\theta)$, the second from applying the definitions of $\delta^B_i$, $\beta_B$  and $U_B$, and the third from canceling terms, and the last from the definition of $\theta_F$.

Since $\theta_F$ was chosen arbitrarily, we have shown $f(f(\theta)) = \theta, \forall \theta$ and thus $f^{-1} = f(\theta)$.

\begin{lemma}
Let $\theta=\{\delta, \beta, T, C\}$ denote model parameters of the factorized forward-dependent CDM ranking model, and let 
$$g(\theta) = \left\{\Bigl\{\delta_i+t_{i}^T\sum_{j\in\mathcal{U}\setminus i} c_j,~\forall i\Bigr\}, \beta, T, -C\right\}.$$ 
The inverse map is the map itself: $g^{-1} = g(\theta)$
\label{lem:low}
\end{lemma}
\textbf{Proof.}
The proof follows the same form and steps as the previous lemma. If $g(g(\theta)) = \theta, \forall \theta$, then $g^{-1} = g(\theta)$. We show the former to be true. Let $\theta_F := \{\delta^F, \beta^F, T^F, C^F\}$ and let $\theta_B := g(\theta_F) = g(\delta^F, \beta^F, T^F, C^F)$. We have,

\begin{align*}
\theta_B=\left\{\Bigl\{\delta_i^F+t_{i}^{FT}\sum_{j\in\mathcal{U}\setminus i} c_j^F,~\forall i\Bigr\}, \beta^F, T^F, -C^F\right\}=\{\delta^B, \beta^B, T^B, C^B\},\\
\end{align*}
where 
\begin{align*}
&\delta^B_i := \delta_i^F+t_{i}^{FT}\sum_{j\in\mathcal{U}\setminus i} c_j^F,~\forall i \\
&\beta_B := \beta_F \\
&T^B := T^F \\
&C^B := -C^F
\end{align*}
Now,
\begin{align*}
g(g(\theta_F)) = g(\theta_B)=g(\{\delta^B, \beta^B, T^B, C^B\}) &= 
\left\{\Bigl\{\delta_i^B+t_{i}^{BT}\sum_{j\in\mathcal{U}\setminus i} c_j^B,~\forall i\Bigr\}, \beta^B, T^B, -C^B\right\} \\
&= 
\left\{\Bigl\{\delta_i^F+t_{i}^{FT}\sum_{j\in\mathcal{U}\setminus i} c_j^F+t_{i}^{FT}\sum_{j\in\mathcal{U}\setminus i} -c_j^F,~\forall i\Bigr\}, \beta^F, T^F, --C^F\right\} \\
&=\left\{\Bigl\{\delta^F_i,~\forall i\Bigr\}, \beta^F, T^F, C^F\right\}.\\
&=\{\delta^F, \beta^F, T^F, C^F\} = \theta_F.
\end{align*}
where the first line follows from applying the definition of $g(\theta)$, the second from applying the definitions of $\delta^B_i$, $\beta_B$  and $T_B$ and $C_B$, and the third from canceling terms, and the last from the definition of $\theta_F$.

Since $\theta_F$ was chosen arbitrarily, we have shown $g(g(\theta)) = \theta, \forall \theta$ and thus $g^{-1} = g(\theta)$.

\newpage
\section{Hyperparameter tuning}
\label{sec:hyper}

The models in this work each require the selection of various hyperparameters; the low-rank CDM requires the selection of the embedding dimension $r$, a stratified model is specified by $K$ and $\lambda_{\mathcal{L}}$, the number of strata and amount of Laplacian regularization, respectively, and all models apply non-zero $\ell_2$ regularization, $\lambda>0$. We tune these hyperparameters via 5-fold cross validation within our training dataset, selecting the values that minimize validation loss.

\begin{figure*}[h!]
    \centering
    \begin{subfigure}[ht]{0.47\textwidth}
        \centering
        \includegraphics[width=\textwidth]{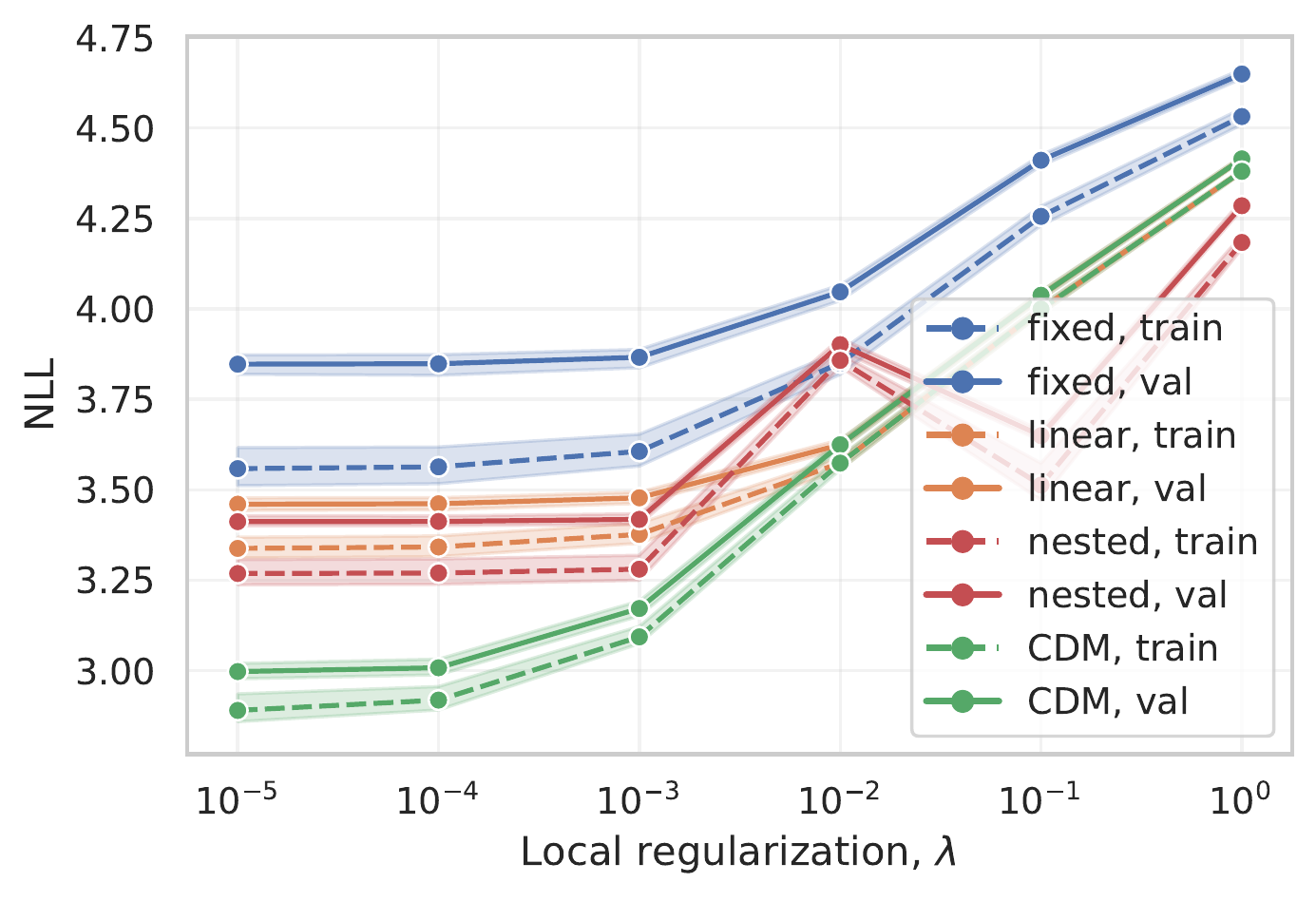}
    \end{subfigure}%
    \hspace{0.5cm}
    \begin{subfigure}[ht]{0.47\textwidth}
        \centering
        \includegraphics[width=\textwidth]{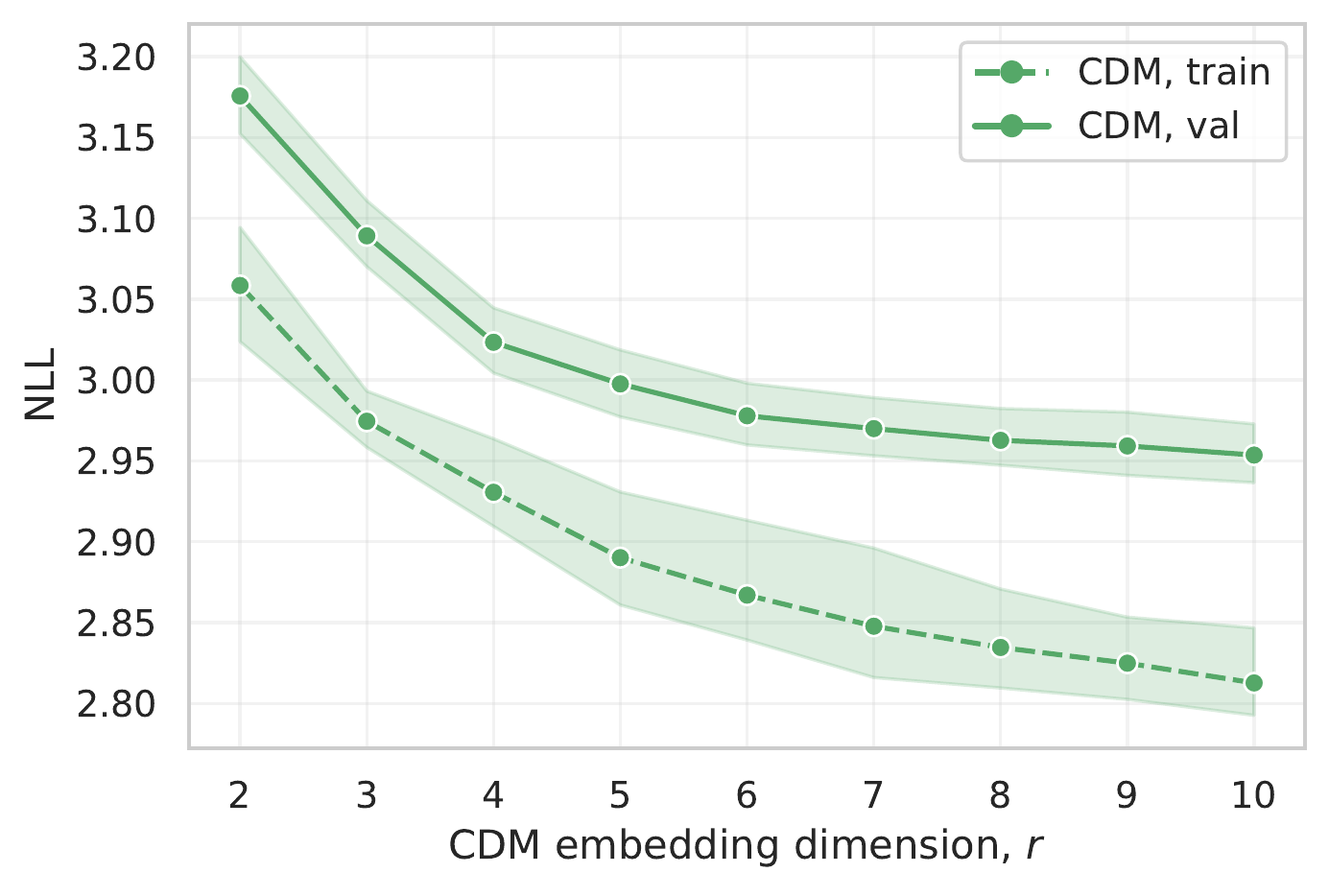}
    \end{subfigure}
    \caption{Left: Amount of local $\ell_2$ regularization, $\lambda$, tuning, used on all model parameters. Band denotes 95\% confidence intervals across 5-folds. We take $\lambda=10^{-5}$ to achieve identifiability of the models while minimizing validation loss. Right: Embedding dimension of the low-rank CDM, $r$, tuning, used for both stratified and non-stratified models. $r=10$ minimizes validation loss.}
    \label{fig:hyper}
\end{figure*}

A minimal amount of local regularization, $\lambda=10^{-5}$, is applied, only towards achieve identifiability of the parameters, and the embedding dimension of the low-rank CDM is selected to be $r=10$ (Figure~\ref{fig:hyper}).

\begin{figure*}[ht]
    \centering
    \includegraphics[width=\textwidth]{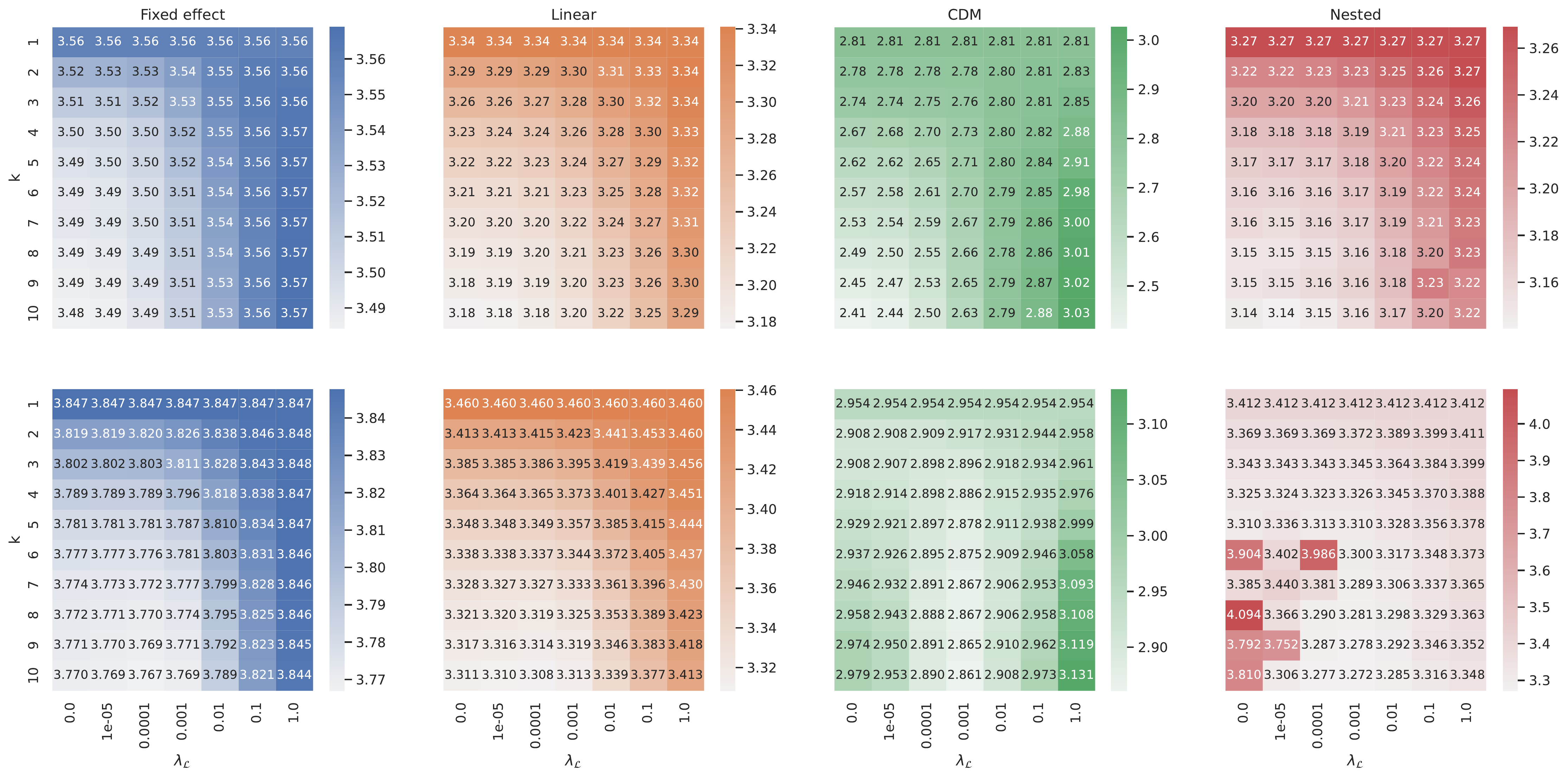}
    \caption{Hyperparameter tuning for the number of stratification buckets, $K$, and amount of stratified regularization, $\lambda_\mathcal{L}$ for the three main models + nested. Top row denotes training loss, bottom row shows validation. $(k,\lambda_\mathcal{L})=[(10, 10^{-4}), (10, 10^{-4}), (10, 10^{-3}), (10, 10^{-3})]$ minimizes validation loss for fixed effect MNL, linear MNL, CDM, and nested models, respectively.}
    \label{fig:strat_tuning}
\end{figure*}

In Figure~\ref{fig:strat_tuning}, we see that all training errors (top row) are minimized with the most strata and least regularization, resulting in a model with maximum flexibility to fit the training data.
However in validation (bottom row), the stratified CDM pays a large price with more stratification and less regularization.
The multiplicative increase in parameters by the stratification leads to more significant over-fitting to the training data for the CDM than the linear and fixed-effect models, so more regularization is needed in this case.
Thus, the selected stratification hyperparameters, ($K$, $\lambda_{\mathcal{L}}$), are ($10$, $10^{-4}$) for the fixed-effect MNL and linear MNL models, and ($10$, $10^{-3}$) for the CDM and nested models.
We summarize the tuned model hyperparameters in Table~\ref{tab:hyperparams}.

\begin{table}[ht]
    \centering
    \caption{Tuned model hyperparameters.}
    \begin{tabular}{c|c||c|c|c|c}
        \multirow{2}{*}{\textbf{Hyperparameter}}  & \multirow{2}{*}{\textbf{Applies to}}   & \multicolumn{4}{c}{\textbf{Value}}     \\\cline{3-6}
                        &           & Fixed-effect    & Linear    & CDM    & Nested\\\hline\hline
        $\lambda$       & All        & \multicolumn{4}{c}{$10^{-5}$}\\\hline
        $r$             & CDM       &   - &  - & $10$  & -\\\hline
        $K$             & \multirow{2}{*}{Stratified} & 10   & 10   & 10 & 10\\
        $\lambda_L$     &           & $10^{-4}$ & $10^{-4}$ & $10^{-3}$ & $10^{-3}$\\
    \end{tabular}
    \label{tab:hyperparams}
\end{table}

\clearpage

\section{Parameter estimates}
\label{sec:params}
\subsection{Stratified fixed-effects, $\hat\delta$}
\begin{figure}[h!]
    \centering
    \includegraphics[width=\textwidth]{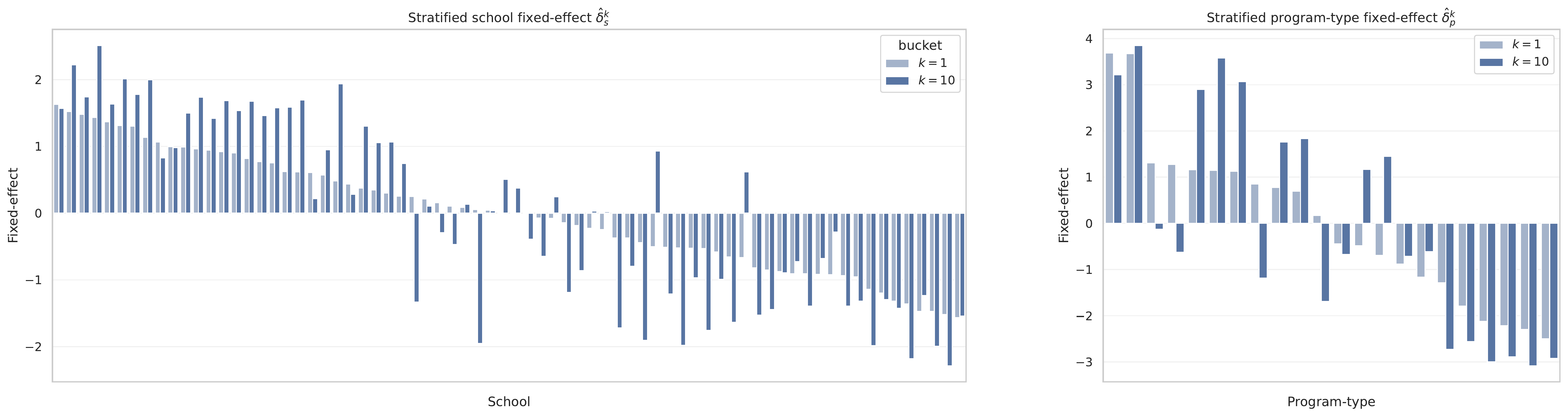}
    \caption{
    Stratified fixed-effects for $k=\{1,10\}$. Schools (left) and program-types (right) are sorted on the $x$-axes according to $k=1$ fixed-effects. Later distributions shift away from top-choice-popular programs.}
    \label{fig:strat_logits}
\end{figure}

In Figure~\ref{fig:strat_logits}, we plot the fixed-effects learned by the stratified fixed-effect MNL at ranks $k=\{1,10\}$. We sort the schools and program-types on the $x$-axes by the $k=1$ model's parameter estimates, $\hat \delta^1$. We see that the later distributions shift weight away from the top-choice-popular alternatives. This redistribution yields improved performance for the fixed-effect model down-rank. See Figure~\ref{fig:completion} for evidence of this result.

\subsection{Stratified vs.\ non-stratified $\hat\beta$}
\begin{figure}[ht]
    \centering
    \includegraphics[width=0.75\textwidth]{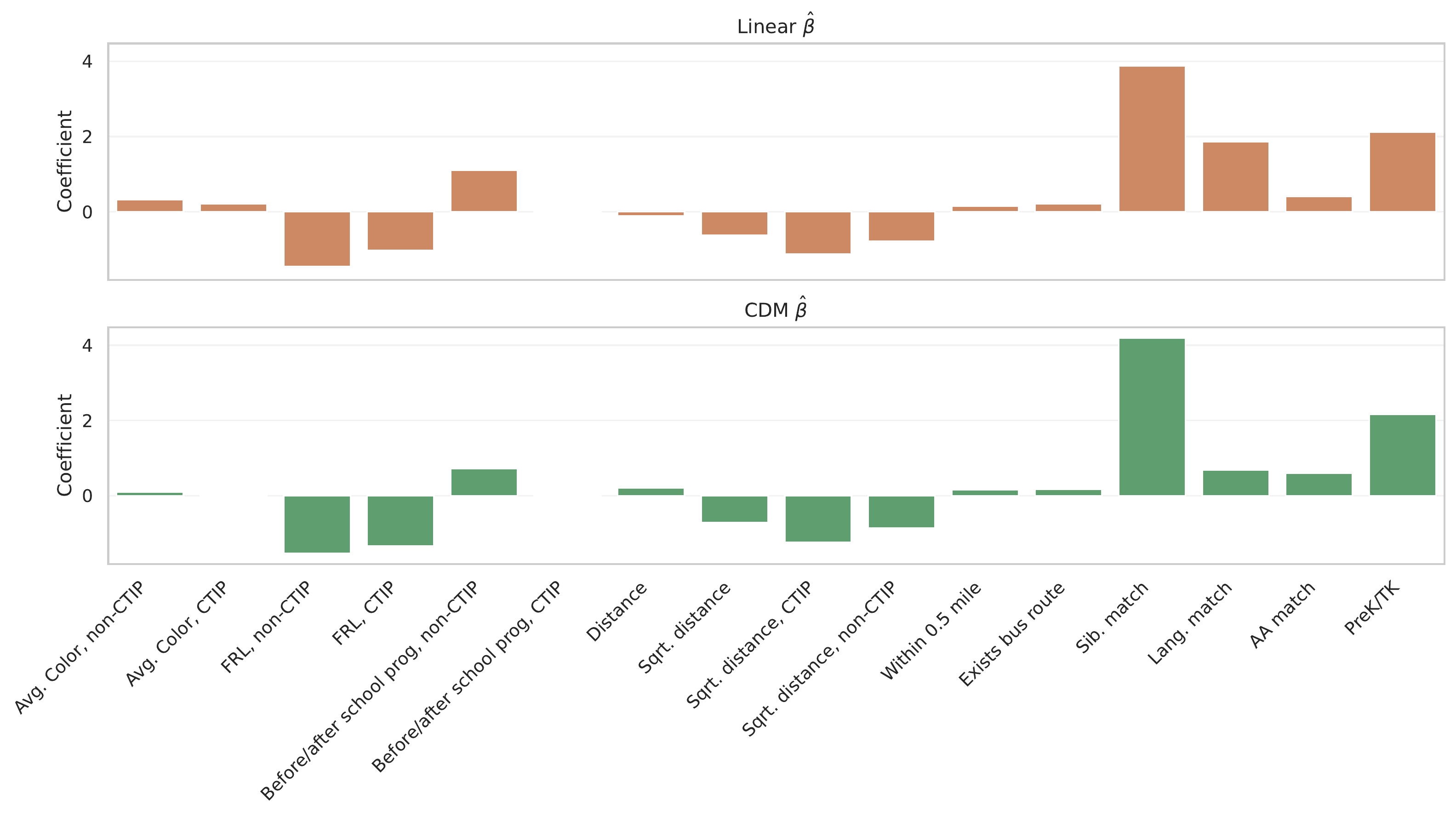}
    \caption{Learned linear weights, $\hat \beta$, from training the linear MNL and CDM models on 2017-18 school year preference data. The sign of most coefficients align with intuition in both linear MNL and CDM models. For example, the coefficients on (square-root) distance is negative, signaling that there is a preference for proximity to home. Meanwhile the parameters for before/after school programs, PreK/TK continuation, language match, and sibling match are all positive.
    }
    \label{fig:beta}
\end{figure}

\begin{figure}
    \centering
    \includegraphics[width=\textwidth]{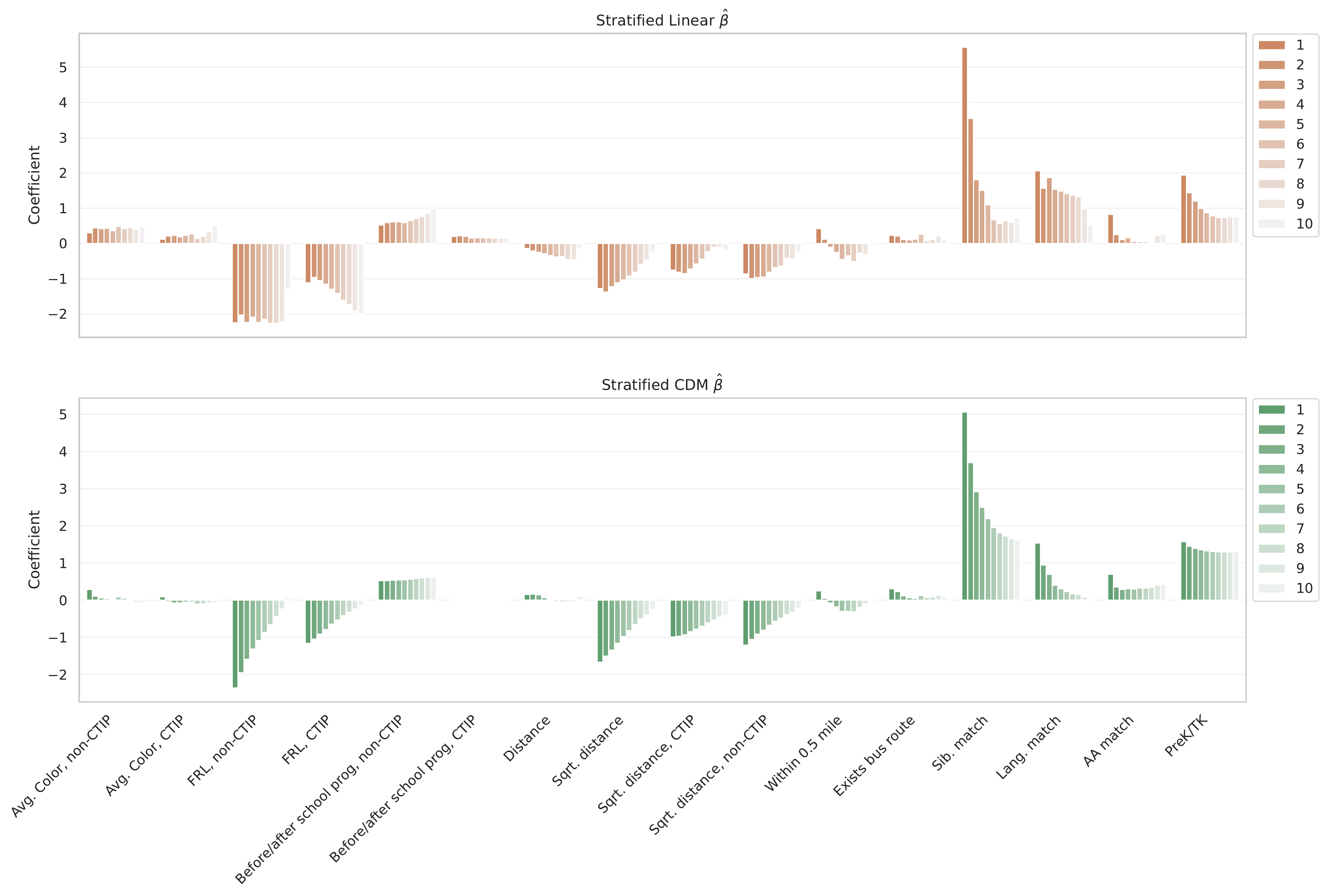}
    \caption{Learned linear model estimates $\hat \beta_i$ from training the stratified linear MNL and CDM models on 2017-18 school year preference data. Previous rank-heterogeneous models \cite{Hausman1987} assume that coefficient magnitudes contract towards zero down-rank, but relaxing this assumption we find coefficients frequently show non-linear/non-monotonic behaviors down rank.}
    \label{fig:stratbeta}
\end{figure}

Next we report the coefficient estimates $\hat \beta$ from the non-stratified (Figure~\ref{fig:beta}) and stratified (Figure~\ref{fig:stratbeta}) linear MNL and CDM models.
In Figure~\ref{fig:beta}, the sign of most coefficients align with intuition in both linear MNL and CDM models. 
For example, the coefficients on distance are negative, signaling that there is a preference for proximity to home. Meanwhile the parameters for before/after school programs, PreK/TK continuation, language match, and sibling match are all positive. 

When stratifying the models into $K=10$ strata for both the linear and CDM models respectively, we see in Figure~\ref{fig:stratbeta} that the parameter magnitudes mostly diminish towards zero as we model down-rank choices.
This contraction is consistent with either a less confident model--the choice datasets shrink in size as we subset on lower rank choices as the number of families ranking at least $k$ alternatives decreases with increasing $k$--or less confident assembly of down-rank preferences by households. The latter is the main hypothesis of Hausman and Ruud~\cite{Hausman1987}; these authors developed a heteroscedastic model with uniformly diminishing coefficients at each rank position to model this effect.

However, there are a few examples of non-monotonically shrinking coefficients in the stratified linear model's coefficients.
Most notably, the coefficients on the fraction eligible for reduced lunch actually becomes more negative down rank within the CTIP1 and stays constant for the non-CTIP1 populations.
In this way, the regularized stratified model allowed us to learn from truly evolving, not simply vanishing, preferences. 
This finding is an example where Hausman and Ruud's work alone is insufficient in modeling household values.

\subsection{Nested MNL}
\label{sec:nested}
The nested model extends the MNL to allow groups of alternatives, called \emph{nests}, to be ``similar'' to each other in an unobserved way; that is, to have correlated error terms. To define the model, let $K$ be the number of predefined nests. 
Denote by $B_i\subseteq \mathcal{U}$ the set of alternatives assigned to nest $i$ for $i\in[K]$, and $B(j)\in[K]$ to be the unique nest membership of alternative $j$. 
Given these nests and memberships, the nested MNL choice probability is given in closed form by the following formula:
\begin{align*}
P(j|i,S) &= \frac{e^{V_{ij}/\lambda_{B(j)}}\left(\sum_{k\in B(j)}e^{V_{ik}/\lambda_{B(j)}}\right)^{\lambda_{B(j)}-1}}{\sum_{\ell=1}^K \left(\sum_{k\in B(\ell)}e^{V_{ik}/\lambda_\ell}\right)^{\lambda_\ell}}\\
&=\frac{e^{V_{ij}/\lambda_{B(j)}}}{\sum_{k\in B(j)}e^{V_{ik}/\lambda_{B(j)}}}\cdot\frac{\left(\sum_{k\in B(j)}e^{V_{ik}/\lambda_{B(j)}}\right)^{\lambda_{B(j)}}}{\sum_{\ell=1}^K \left(\sum_{k\in B(\ell)}e^{V_{ik}/\lambda_\ell}\right)^{\lambda_\ell}}\\
&=P_{j|i,B(j)}\cdot P_{B(j)|i},
\end{align*}
where $\lambda_i$ is a measure of independence in nest $i$. When $\lambda=1$, the model is identical to the standard MNL and nests are abandoned, and $\lambda<1$ indicates positive correlation amongst nest alternatives.

We implement a nested MNL in our setting by nesting the program offerings by program type. Specifically, by `Chinese Language', `Filipino Language', `General Education', `Japanese Language', `Korean Language', `Spanish Language', and `Special Education' offerings, for a total of $K=7$ nests spanning the full menu of available programs in SFUSD. Representative utilities are taken to be identical to the linear MNL specification in Eq.~\eqref{eq:lin} with the same covariates. 
As with the fixed-effect, linear MNL, and CDM models, we run Adam with default parameters, adding $\ell_2$ regularization in accordance with our strength selection of $1e-5$ in Table~\ref{tab:hyperparams}. 
Model parameters are updated over batches of training data until reaching $\texttt{max\_epoch}=1000$ or convergence, i.e., when the absolute difference in losses is less than $\epsilon=1e-4$. 
Learned scale parameters $\lambda_k\in \mathbb{R}^K$ are given in Table~\ref{tab:lambda}.
\begin{table}[ht]
    \centering
    \caption{Learned independence parameters, $\hat \lambda_k$, of our nested MNL model. Lowest $\hat \lambda_k$, and therefore highest within-nest correlation, in \textbf{bold}.}
    \begin{tabular}{l|c||c}
        Nest, $B_k$             & Nest size, $|B_{k}|$ & Parameter, $\hat\lambda_k$ \\\hline
        General Education       & 65        & 0.4271\\
        Spanish Language        & 32        & 0.6562\\
        Special Education       & 27        & \bf{0.3634} \\
        Chinese Language        & 24        & 0.7709 \\
        Korean Language         & 2         & 0.5118 \\
        Filipino Language       & 2         & 0.5549 \\
        Japanese Language       & 2         & 0.5708 
    \end{tabular}
    \label{tab:lambda}
\end{table}

In Figure~\ref{fig:overall}, we find that the nested MNL performs almost identically to the uncorrelated linear model, with only marginal performance improvement. 
The CDM model outperforms even the nested model as it learns a more nuanced similarity amongst the alternatives, and does so implicitly rather than through explicitly defined subsets.

\begin{figure}
    \centering
    \includegraphics[width=\textwidth]{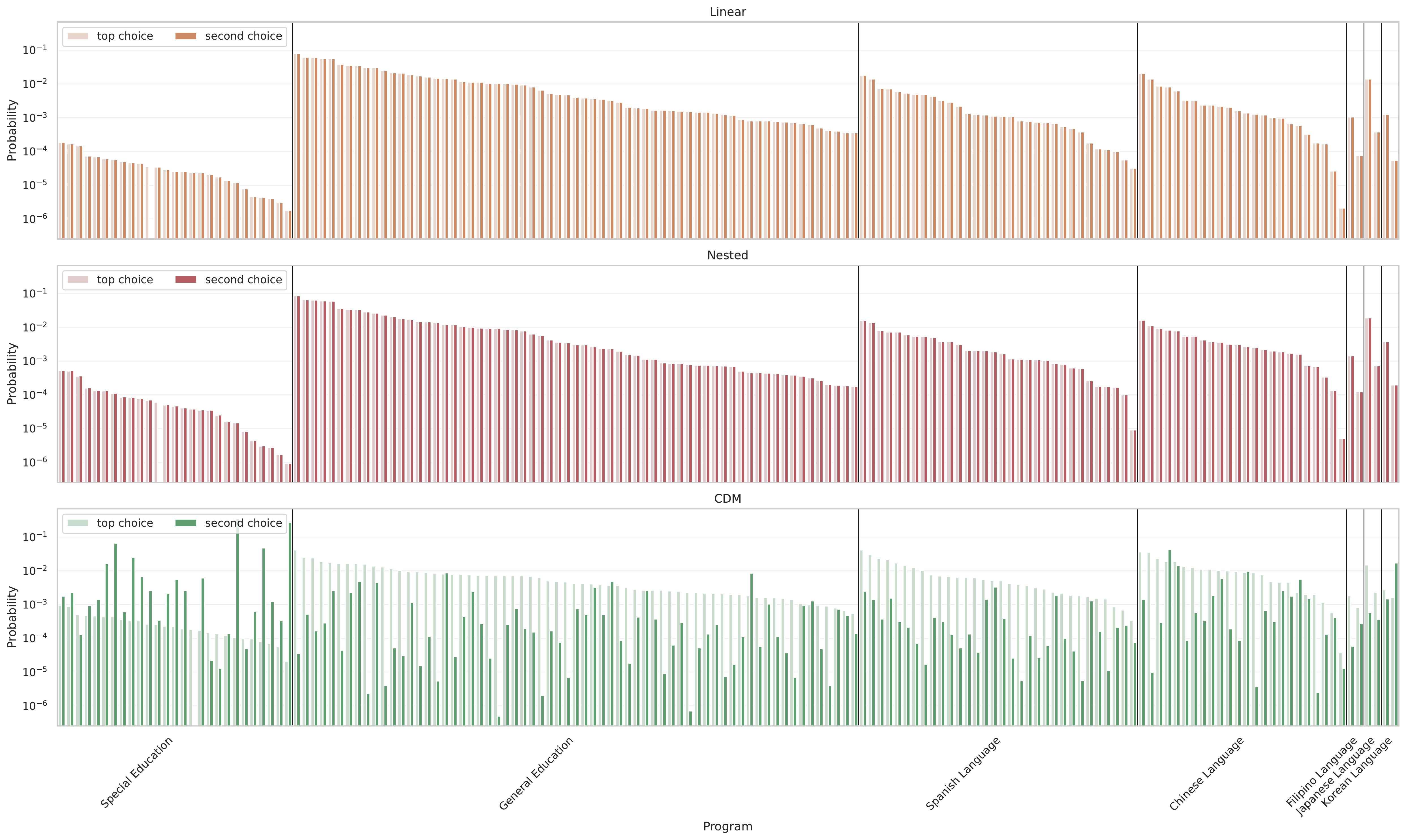}
    \caption{Linear MNL, nested MNL, and CDM choice probabilities in top-choice (no context) and second-choice (one-chosen program) across all available alternatives for an example household \emph{that chose a special education program first}. In the context of the selected alternative, linear and nested MNL models do not significantly redistribute second-choice probabilities, whereas the CDM distribution is more adaptive to the information of chosen alternatives.}
    \label{fig:nested_probs}
\end{figure}

To investigate this effect further, we plot top- and second-choice probabilities of the linear, nested, and CDM models for a select household in the test data in Figure~\ref{fig:nested_probs}. Specifically, this household selected a special education program for their student in their top-position. One would expect both the nested MNL and CDM models to make use of this information in their second choice distribution. However, the 1st and 2nd probability distributions are effectively identical in the nested model. The nested model ``redistributes'' the selected choice's first-choice probability to the remaining programs in a way that favors other special education offerings, but the effect is minimal as the selection of any special education program first is relatively rare in our data. The CDM, on the other hand, makes great use of this intel and dramatically increases the likelihood of choosing other special education programs. The CDM is capable of modeling behavioral signals that are not present in household or program covariates, and therefore presents measurable advantages over the rank-homogeneous models studied in this work.

\clearpage

\section{Model Accuracy by Subpopulation}
\label{sec:subpop}
In addition to reporting goodness of fit and overall accuracy of the models, it is important to also evaluate model performances by
(1) their ability to predict the choice of easy-to-predict subgroups (e.g. sibling), serving as a sanity check, and (2) their ability to predict the choices of subgroups of interest to the decision-maker (e.g. Black, Hispanic/Latino, CTIP1), since the choice model will eventually be used to predict outcomes of policies on these subgroups. We drop the null and fixed effect models from these plots as they are relatively inaccurate, evidenced in Figure~\ref{fig:completion}.

\begin{figure*}[h!]
    \centering
    \begin{subfigure}[ht]{\textwidth}
        \centering
        \includegraphics[width=0.74\textwidth]{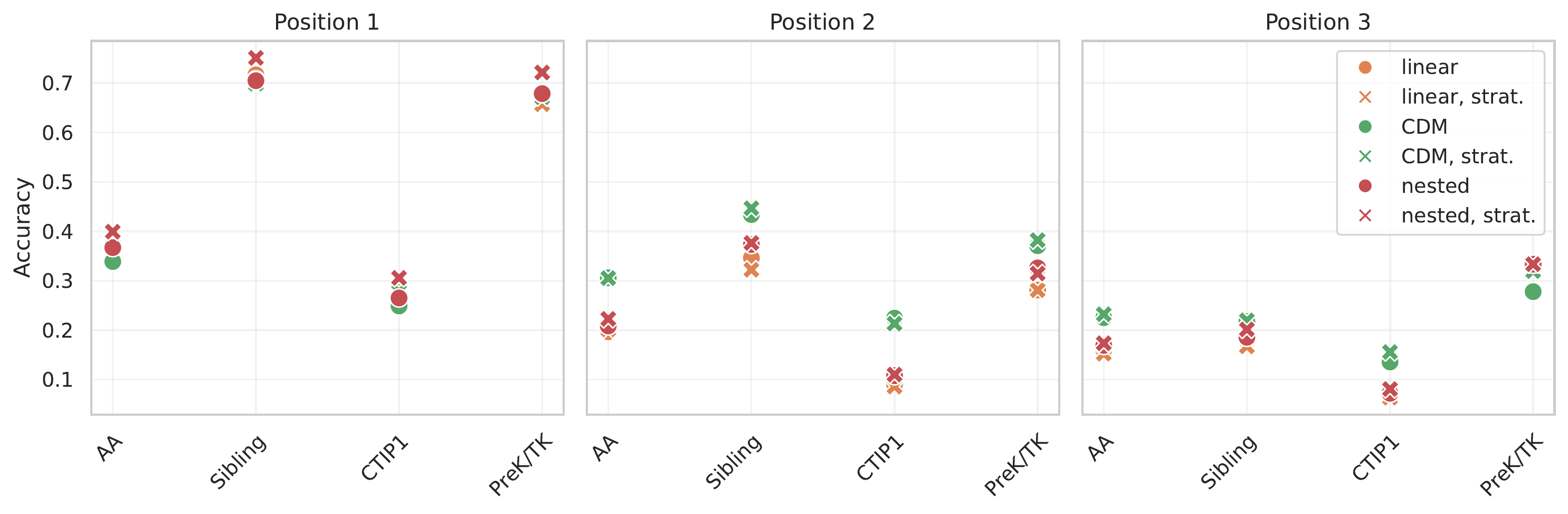}
        \caption{Accuracy by priority categories. See Section~\ref{sec:sfusd} for definitions of all priority categories. }
        \label{fig:by_subpop_priority}
    \end{subfigure}\\
    \vspace{0.25cm}
    \begin{subfigure}[ht]{\textwidth}
        \centering
        \includegraphics[width=0.74\textwidth]{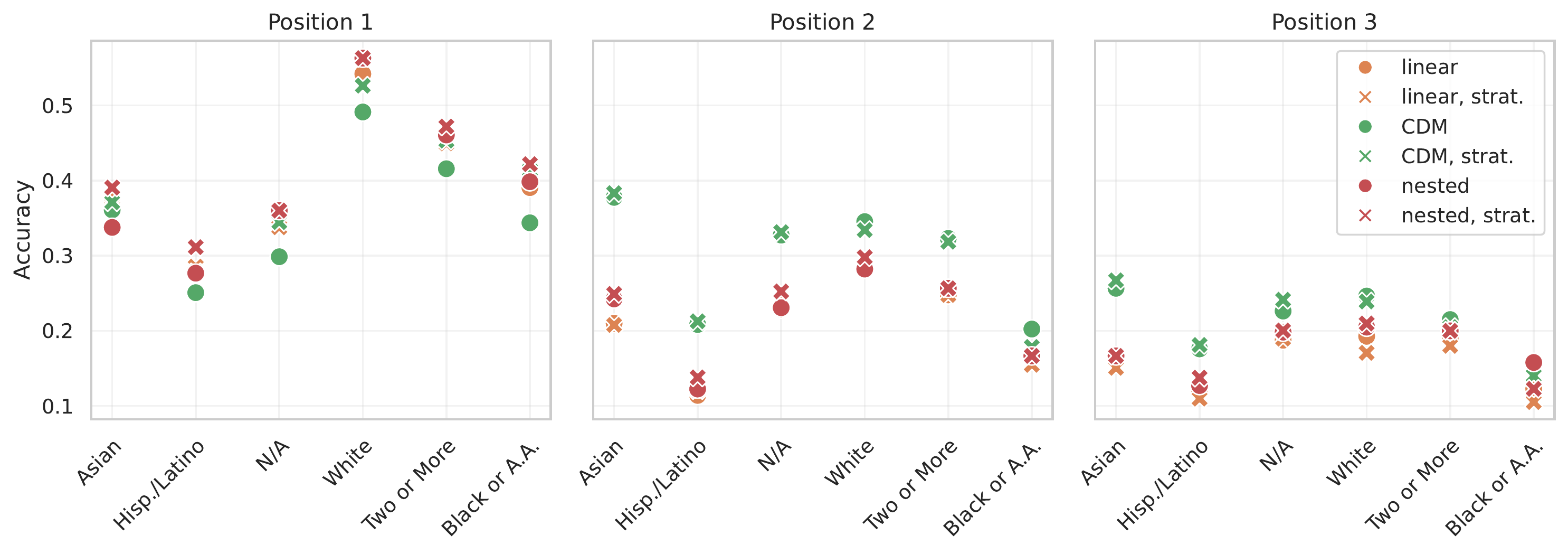}
        \caption{Accuracy by race/ethnic group. }
        \label{fig:by_subpop_race}
    \end{subfigure}\\
    \vspace{0.25cm}
    \begin{subfigure}[ht]{\textwidth}
        \centering
        \includegraphics[width=0.74\textwidth]{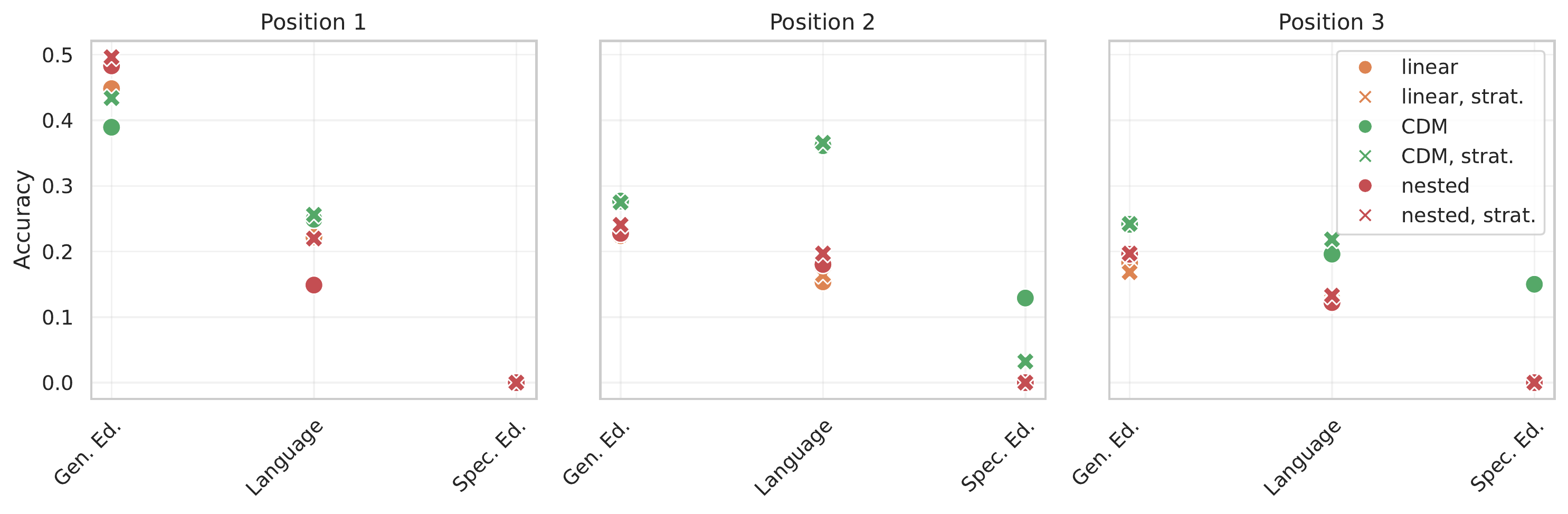}
        \caption{Accuracy by which program type was ranked in the first position.}
        \label{fig:by_subpop_prog}
    \end{subfigure}
    \caption{Prediction Accuracy at $k=[1,2,3]$ over key sub-populations. First column corresponds to top-choice prediction ($k=1$), last is third-choice ($k=3$). Groups are ordered largest to smallest from left to right on x-axes.
    }
    \label{fig:by_subpop}
\end{figure*}

In Figure~\ref{fig:by_subpop_priority}, we see that sibling and PreK/TK priority groups show highest accuracy in top choice, as these groups gain strong priority to specific schools and tend to rank those schools first. The models systematically under perform in accuracy on CTIP1 (Figure~\ref{fig:by_subpop_priority}), Hispanic/Latino, and Black/African-American populations (Figure~\ref{fig:by_subpop_race}). These groups are either highly varied in their demonstrated preferences, making them harder to predict for, or there was not enough training data present for these populations. Finally, the CDM demonstrates the largest lead in predicting second and third choices for households who ranked a special education or language program first (Figure~\ref{fig:by_subpop_prog}).
\clearpage 

\section{Model consistency}
\label{sec:consistency}
Here we note the sampling consistency--how similar sampled choices are--via two metrics: weighted Kendall's $\tau$ correlations amongst generated lists per household, and sampling consistency when completing at position $k$.
We first present weighted Kendall's tau correlations between generated preferences and then report how often the model agrees with itself when predicting the $k$-th choice when given the true first $k-1$ choices, $R_{i,k-1}$.

\begin{figure*}[h!]
    \centering
    \begin{subfigure}[ht]{0.47\textwidth}
        \centering
        \includegraphics[width=\textwidth]{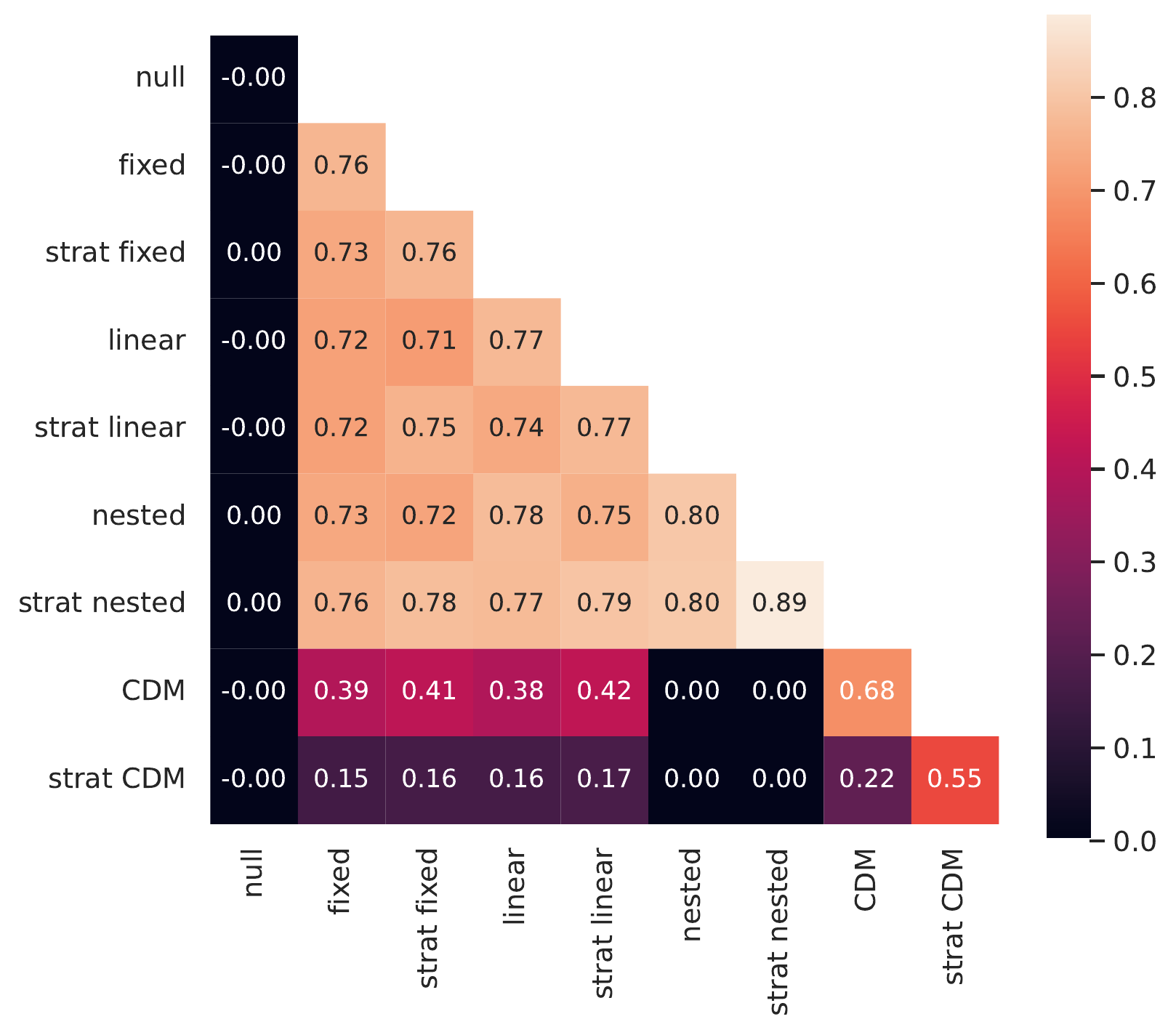}
    \end{subfigure}%
    \hspace{0.75cm}
    \begin{subfigure}[ht]{0.48\textwidth}
        \centering
        \includegraphics[width=\textwidth]{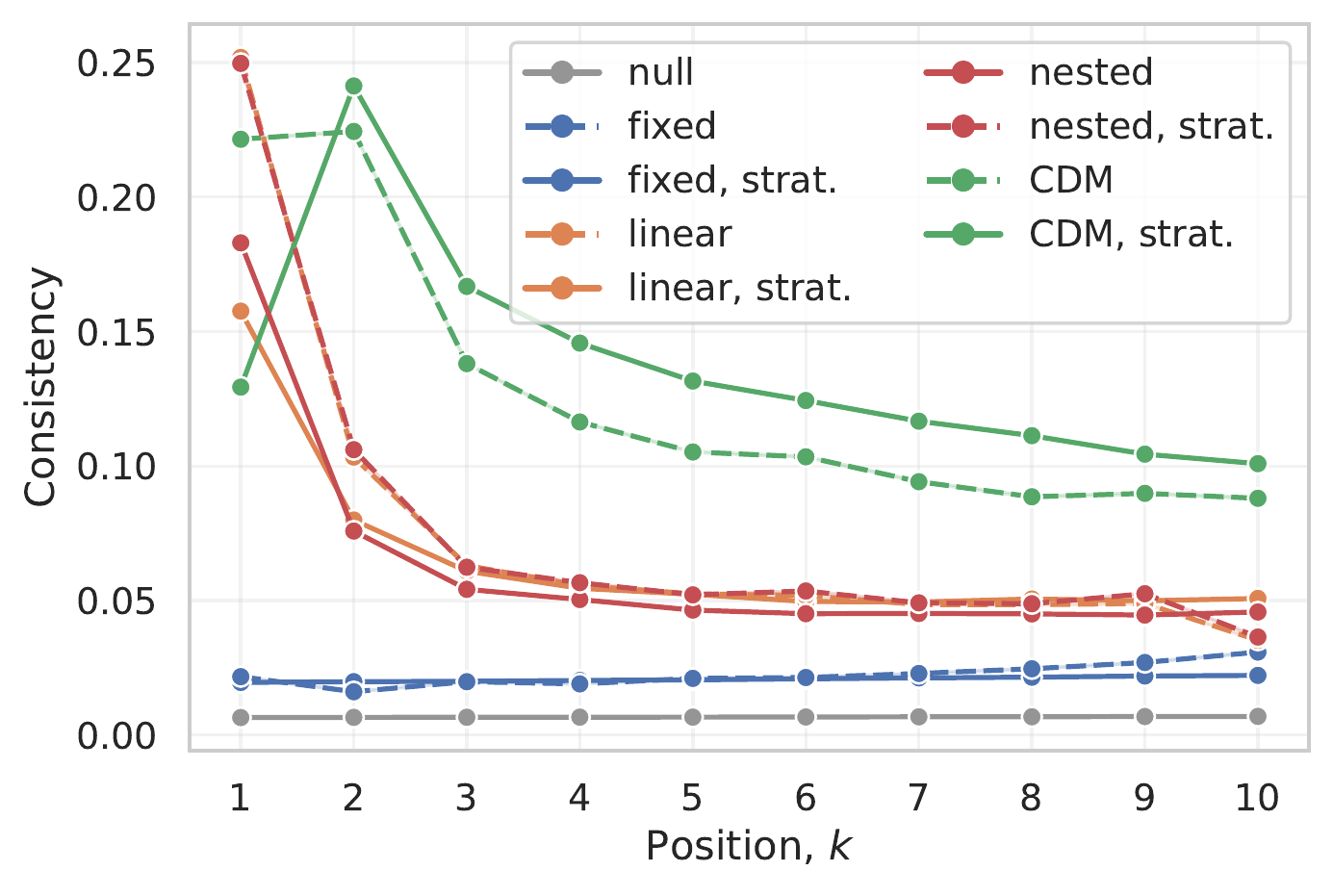}
    \end{subfigure}
    \caption{Correlation and consistency statistics for the 8 studied models. Left: average weighted Kendall's $\tau$ correlations between simulated preferences. Preferences generated by the null distribution are unlike the rest, as expected, and those by the CDM models are more unlike fixed-effect, linear, and nested samples than they are unlike each other. Right: Consistency vs rank position for the studied models. Null and fixed-effect remain fairly (in)consistent, whereas the nested, linear, and CDM models report more consistent predictions in top predictions. The CDM is most consistent from position 2 onward due to the information of the top-choice.}
    \label{fig:consistency}
\end{figure*}

To compute the Kendall's $\tau$ statistic, we sample choices sequentially from each model and generate $N=100$ full rankings, total orderings of the $m$ elements of $\mathcal{U}$, for each household $i$.
Figure~\ref{fig:consistency} shows the weighted Kendall's $\tau$ correlation between these generated preferences across all model pairs, averaged over all students. 
As expected, the null model generates preferences that are completely uncorrelated with itself and the rest. 
Unsurprisingly, the CDM model class generates preference samples that are more unlike the other samples, as seen in the two CDM rows. The CDM models are susceptible to a snow-ball effect when generating full preferences from scratch--the top choices have strong down-stream effects on later choices whereas the linear and fixed-effect (and effectively nested, in this case) MNL models are identically and effectively independently sampled down rank, resulting in more similar lists to themselves and each other.

The right plot of Figure~\ref{fig:consistency} displays the consistency of the model predictions at the $k$-th position when given the true first $k-1$ choices made. 
To measure consistency, we make $N=100$ $k$-th choice predictions for student $i$, compute the fraction of $N \choose 2$ pairs that agree with each other, and average these fractions over all students $i$ in the test set.
The null and fixed-effect classes of models remain similarly (in)consistent throughout, indicating similar probability distributions across the ranking, regardless of prior choices.
The nested, linear, and CDM models predict very consistently--more pairs of predictions at $k$ agree with one another--in the top few choices as they learn from priority statuses and program affiliations.
The CDM shows strong consistency in position 2--the introduction of the first context effect skews the distribution per household to be more modal in follow up program selections.
The effect gradually diminishes after the second rank position, however, as the context effects get averaged out by the growth of the context set, Eq.~\eqref{eq:cdm}.

\end{document}